\documentclass{article}

\usepackage{graphicx,amsmath,amssymb}
\usepackage{eclbkbox}
\usepackage{ifpdf}
\ifpdf
\usepackage[%
  pdftitle={Construction of Reversible Lattice Molecular Automata},%
  pdfauthor={Takayuki Nozawa},%
  pdfsubject={The preprint document class elsart},%
  pdfkeywords={Reversible Cellular Automata, Molecular Aggregation,
  Self-organization, and Artificial Chemistry},%
  pdfstartview=FitH,%
  bookmarks=true,%
  bookmarksopen=true,%
  breaklinks=true,%
  colorlinks=true,%
  linkcolor=blue,anchorcolor=blue,%
  citecolor=blue,filecolor=blue,%
  menucolor=blue,pagecolor=blue,%
  urlcolor=blue]{hyperref}
\else
\usepackage[%
  breaklinks=true,%
  colorlinks=true,%
  linkcolor=blue,anchorcolor=blue,%
  citecolor=blue,filecolor=blue,%
  menucolor=blue,pagecolor=blue,%
  urlcolor=blue]{hyperref}
\fi

\newcommand{\NEWTERM}[1]{\emph{#1}}
\newcommand{\STRESS}[1]{\emph{#1}}

\newcommand{\VECTOR}[1]{{\boldsymbol #1}}

\newcommand{\Z}{\mathbb{Z}}

\newcommand{\NEIGHBOUR}{\mathcal{N}}
\newcommand{\SIGN}{\mathrm{sgn}}
\newcommand{\PARITY}{\mathrm{parity}}
\newcommand{\MType}{\mathit{mt}}
\newcommand{\MOri}{\mathit{mo}}
\newcommand{\TKE}{\mathit{tke}}
\newcommand{\RKE}{\mathit{rke}}
\newcommand{\MBond}{\mathit{mb}}
\newcommand{\Heat}{\mathit{h}}
\newcommand{\PDir}{\mathit{pd}}
\newcommand{\HeatMax}{H_{\mathrm{max}}}
\newcommand{\VACUUM}{\mathrm{V}}
\newcommand{\WATER}{\mathrm{W}}
\newcommand{\PHILIC}{\mathrm{I}}
\newcommand{\PHOBIC}{\mathrm{O}}
\newcommand{\AMPHIPHILE}{\mathrm{A}}
\newcommand{\DirRotation}{\Delta}
\newcommand{\PermMTC}{\psi_{\mathrm{mtc}}}
\newcommand{\BondReadjustment}{\psi_{\mathrm{br}}}
\newcommand{\PermMolHeat}{\psi_{\mathrm{th}}}%
\newcommand{\PotentialChangeByMolRemoval}{{\delta^{(-)}V}}
\newcommand{\PotentialChangeByMolPutting}{{\delta^{(+)}V}}
\newcommand{\PotentialChangeCompensationByHeats}{\psi_{\mathrm{pcc}}}
\newcommand{\PermRot}{\rho}
\newcommand{\PotentialChangeByRotation}[2]{{\delta(#1, \DirRotation^{#2})}V}
\newcommand{\HeatShift}{\sigma_{\mathrm{h}}}
\newcommand{\HeatCollision}{\phi_{\mathrm{h}}}
\newcommand{\AltPermRot}{\rho_\mathrm{a}}

\begin{document}


\title{Construction of Reversible Lattice Molecular Automata}



\author{Takayuki Nozawa and Toshiyuki Kondo}
\date{tknozawa@cc.tuat.ac.jp}


\maketitle

\begin{abstract}

Several cellular automata (CA) models have been developed to simulate
 self-organization of 
multiple levels of structures.
However, they do not obey microscopic reversibility and conservation
laws.
%
In this paper, we describe the construction of a \NEWTERM{reversible
lattice molecular automata} (RLMA) model, which simulates molecular
interaction and self-organization of higher-order structures.
%
The model's strict reversibility entails physically relevant
conservation laws, and thus opens a way to precise application and
validation of the methods from statistical physics in studying
the necessary conditions for such multiple levels of self-organization.

\vspace{1em}
\end{abstract}


\section{Introduction}

Possessing and utilizing multiple levels of self-organized
structures---sometimes addressed as \STRESS{dynamical
hierarchies}\cite{Rasmussen2001,McGregor2005}---is a characteristic
feature of biological systems.
Cellular automata (CA) and similar discrete paradigms have been
effective in modeling such dynamical self-organization hierarchies.
In the context of molecular aggregation, \NEWTERM{lattice molecular
automata} (LMA) simulates self-organization of water (polar solvent),
monomers, and polymers into clusters and higher-order structures such as
micelles\cite{Mayer1997,Mayer1998,Mayer2000}, and similar models have
been developed to simulate organization of compartment structure and
proto-cell-like self-reproduction\cite{Ono2005,Hutton2007}.

However, these models do not obey microscopic \STRESS{reversibility} and
conservation laws, and therefore, the possibility and stability of the
self-organized structures in these models are, to some extent, implied
in their irreversible time evolution rules.
Under the laws of physics, stable persistence of an organized structure
requires effective utilization of limited resources and smooth disposal
of generated entropy. Therefore, the constraint of reversibility should
not be omitted in studying the necessary conditions for stable
structures, using, for example, the canonical methods of statistical
physics.

In this paper, we describe the construction of \NEWTERM{reversible lattice
molecular automata} (RLMA), which simulates self-organization of water,
monomers, and polymers with a strictly reversible dynamics and
physically appropriate conservation laws.
%
Although several reversible CA models have been proposed to
simulate self-organization processes\cite{Creutz1986,D'Souza1999}, our
model
can simulate self-organization of structures with \STRESS{mobility},
which will be essential for realizing higher-order structures and higher
functionality such as autonomy.

The rest of the 
paper is organized as follows. 
A formal definition of CA is provided and useful techniques in constructing
reversible CA are briefly reviewed in Section 2.
%
Features of the original LMA model and its relation to other models are
also briefly reviewed there.
The construction of our RLMA model is described in Section 3, along with the
conservation laws derived from the reversible dynamics.
Some simulations of monomers and polymers in polar solvent are presented
in Section 4.
Finally, our conclusion is drawn in Section 5.
Appendix A presents an alternative approach for implementing
reversible molecular rotation.

\section{Preliminaries}\label{sec:preliminaries}

\subsection{Formalization of CA}\label{subsec:CA_formalization}

On a $d$-dimensional spatial lattice $\Z^d$, each site (cell) $\VECTOR{i}
\in \Z^d$ is assigned with its \NEWTERM {local state} $\omega_{\VECTOR{i}}
\in A$.  The finite set $A$ of local states is called an 
\NEWTERM{alphabet}.
A specification of local states over the whole space $\omega \triangleq
(\omega_{\VECTOR{i}})_{\VECTOR{i} \in \Z^d} \in \varOmega = A^{\Z^d}$ is
called a \NEWTERM{global state} or \NEWTERM{configuration}.

The dynamics of a CA is given by the \NEWTERM{local transition map}
$\varphi$ as
\begin{equation}
 \label{eq:local_update_rule}
 \omega_{\VECTOR{i}}^{t+1}
 = \varphi \left((\omega_{\VECTOR{j}}^{t})_{\VECTOR{j} \in \NEIGHBOUR(\VECTOR{i})}\right), 
\end{equation}
where the \NEWTERM{neighbor function} $\NEIGHBOUR: \VECTOR{i}
\mapsto (\VECTOR{j}_1, \ldots, \VECTOR{j}_N)$ defines interaction range
for each site $\VECTOR{i}$.
By applying the local map $\varphi$ over the lattice, the
\NEWTERM{global transition map}
\begin{equation}
 \label{eq:global_update_rule}
 \omega^{t+1} = \varPhi(\omega^{t})
\end{equation}
from a configuration at $t$ to the one at $t+1$ is derived.
Although the application of $\varphi$ over the space is synchronous in
simple CA, making it asynchronous can be effective in satisfying
reversibility and other constraints, as shown later.
Furthermore, in more complicated CA, the local map $\varphi$ consists
not of a single map but of several maps (sub-steps), and the local states
$\omega_{\VECTOR{i}} \in A$ also have inner structures such as 
``partitions'' or ``layers.''

When the global transition map $\varPhi$ is bijective, that is, when for
any configuration $\omega^{t+1}$ its pre-image $\omega^{t}$ is unique,
the CA is \NEWTERM{reversible} (or \STRESS{invertible}).

\subsection{Construction of reversible CA}\label{subsec:reversible_maps}

Reversibility entails \STRESS{conservation of information}---differences
in states cannot just appear from or vanish into nowhere.  
Hence the manner in which to prevent information loss is crucial in
constructing reversible CA.

\begin{figure}[th]
 \begin{center}
  \includegraphics[width=0.9\textwidth]{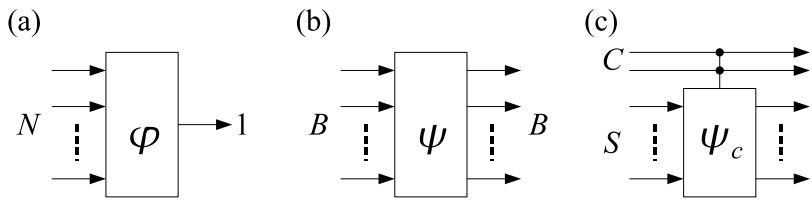}
 \end{center}
 \caption{Mappings to constitute transition: (a) many-to-one mapping,
 (b) permutation, and (c) conditional permutation.}
 \label{fig:UpdateRuleComponentMaps}
\end{figure}

Since the many-to-one local transition map $\varphi$ in
Eq.~(\ref{eq:local_update_rule}) obviously loses information by itself
(Fig.~\ref{fig:UpdateRuleComponentMaps}(a)), the loss should be
prevented by well-counterbalanced distribution among the interacting
cells.  However, designing such maps is far from trivial. Indeed,
judging the reversibility of a global transition map $\varPhi$, given its
corresponding local map $\varphi$, is a difficult task in itself.

An easier method to construct reversible CA is by adopting a
\STRESS{permutation} (reversible transformation on a finite set) $\psi$, 
\begin{equation}
 \label{eq:simple_permutation}
 \psi: A^{B}\rightarrow A^{B}, 
\end{equation}
instead of a many-to-one mapping $\varphi$, as a constituent of the
transition rule (Fig.~\ref{fig:UpdateRuleComponentMaps}(b)).
Here, $B$ denotes a ``block'' of cells under the permutation.  
In \NEWTERM{Partitioning CA} (or \NEWTERM{block CA})\cite{Toffoli1990},
for example, both the reversibility and global transmission of
information are satisfied by combining the permutation and alternation
of different partitioning schemes of a given space into cell-blocks.

The permutation (\ref{eq:simple_permutation}) can be generalized into 
a \NEWTERM{conditional permutation} as
\begin{equation}
 \label{eq:conditional_permutation}
 \psi_{c}: A^{S}\rightarrow A^{S},\quad c \in A^{C}.
\end{equation}
Out of the set $C+S$ of cells that are subject to the mapping, the
states of cells in $C$ work as ``control signals,'' which determine a
permutation for the states of $S$, and reappear unchanged as outputs
(Fig.~\ref{fig:UpdateRuleComponentMaps}(c)).  
The Fredkin gate and the Toffoli gate are well-known examples of conditional
permutations.
%

To prevent information loss, the outputs of the (conditional)
permutations should be reused as inputs or conditional signals, and to
achieve this, one needs to implement certain techniques such as dividing
a time step into several \STRESS{sub-steps}, and arranging the
permutations \STRESS{sparsely and asynchronously in
space-time}\cite{Toffoli2004}.

For synchronous information transmission, one can also use global
\NEWTERM{shifts}, which uniformly displace some partitions or layers of
local states.
The translational movement of free particles can be effectively modeled
by the shifts.
In \NEWTERM{lattice gas automata} (LGA)\cite{Hardy1973,Frisch1986}, for
example, shifts are utilized to express the translation of gas
particles, in combination with permutations that represent the
collisions of the particles.  \NEWTERM{Partitioned CA}\cite{Morita1989}
is virtually equivalent to the LGA.

\subsection{LMA model and other models}
\label{subsec:LMAandLarson}

Various models have been proposed and used to simulate molecular
self-organization processes.  
On the one hand, the molecular dynamics (MD) method models molecules as
particles with appropriate interaction potentials, and solves their
equations of motion in continuous space\cite{Koch1983}.  While the MD
enables microscopically detailed description of the dynamics, size of
the simulated system is restricted by the available computer resources.  
On the other hand, lattice-type models have been successful on
simulating macroscopic behavior of phase separation and aggregation
processes.  Especially the Larson
model\cite{Larson1985,Larson1988,Larson1989} and its
variants\cite{Stauffer1993,Sahimi1994,Liverpool1995,Bernardes1996} are
widely used and many results are reported.
In the traditional Larson model simulating ternary mixture of water, oil
(hydrophobic monomers), and surfactant (amphiphilic polymers), water and
hydrophobic monomers are represented by a set of up and down $(+1, -1)$
spins, respectively, and polymers are represented by strings of spins.
Monte Carlo method is used for update and the ferromagnetic interaction
between the spins induces phase separation, micelle formation, etc.

The original LMA model bridges the gap between the MD method and the
Larson-type models\cite{Mayer2000}: 
While realized in a discrete manner and thus keeping the efficiency of
the lattice setting, it includes some microscopic molecular details,
such as hydrodynamics conserving momenta in the molecular collision,
directions of polar molecules and accompanied anisotropy of molecular
potential energy.
A distinguishing feature of the LMA model is the equienergetic
interaction for the pairs water--hydrophobic monomer and hydrophobic
monomer--hydrophobic monomer, following experimental data on enthalpy
exchanges in mixtures\cite{Privalov1989}.  This setting is in contrast
to the Larson models, which define positive enthalpic gains for oil--oil
interaction but not for water--oil interaction.  Consequently, in the
LMA model phase separation is realized via \STRESS{entropy-driven}
hydrophobic effect, and not enthalpy-driven as in the Larson-type
models.

Although update rule of the LMA model partially keeps the conservation
laws, its dynamics is not microscopically reversible (refer to section
IV.B and V of Ref.~\cite{Mayer1997} for example to see the total energy is
conserved in the mean but not strictly and explicitly).
Therefore, utilizing the techniques introduced in
section~\ref{subsec:reversible_maps}, we construct our RLMA model in the
next section.

\section{RLMA Model}\label{sec:model_construction}

\subsection{Space}\label{subsec:space}

\begin{figure}[th]
 \begin{center}
  \includegraphics[width=0.9\textwidth]{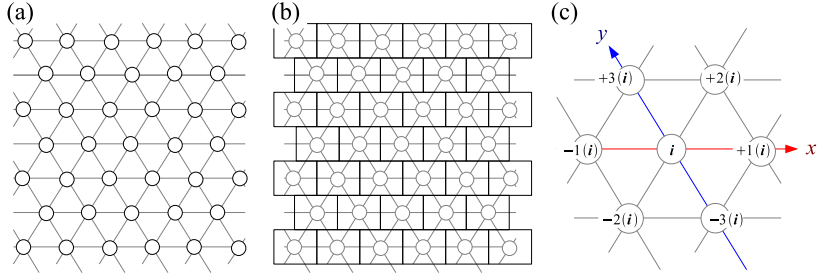}
  \caption{Two-dimensional triangular lattice: (a) structure, (b)
  corresponding cells, and (c) principal directions and nearest neighbors.}
  \label{fig:triangular_lattice}
 \end{center}
\end{figure}

We formalize the RLMA model on the two-dimensional triangular lattice
(Fig.~\ref{fig:triangular_lattice}(a), (b)) as in the
literature\cite{Mayer1997,Mayer1998}, although generalization to other
lattice structures and to higher dimensions will be straightforward.
We use the variable $l \in \{+1, +2, +3, -1, -2, -3\} \equiv L$ to
denote the principal directions, and $l(\VECTOR{i})$ to denote cell
$\VECTOR{i}$'s nearest neighbor in direction $l$, as shown in
Fig.~\ref{fig:triangular_lattice}(c).
$L$ corresponds to $\{0,\pi/6, \ldots,5\pi/6\}$ in the equilateral
triangular lattice with a proper coordinate system
(Fig.~\ref{fig:triangular_lattice}(c)), and on $L$, we define a cyclic
permutation $\DirRotation$ of length 6,
\begin{equation}
 \label{eq:def_dir_rotation_operator}
  \DirRotation = \left(
	    \begin{array}{llllll}
	     +1 & +2 & +3 & -1 & -2 & -3\\
	     +2 & +3 & -1 & -2 & -3 & +1\\
	    \end{array}
	   \right), 
\end{equation}
which corresponds to $+\pi/6$ rotation operator for the principal directions.

\subsection{Local states}\label{subsec:local_states}

Each local state has the \STRESS{layers} (internal data structure) shown
in Table~\ref{tab:structure_local_states}.
\begin{table}[hbt]
 \caption{Structure of local states.}
 {\label{tab:structure_local_states}
  \begin{tabular}{|l|p{40ex}|}
  \hline
   Layer name and variable & Values\\
   \hline\hline
  Molecular type $\MType_\VECTOR{i}$
   & Water ($\WATER$), hydrophilic monomer ($\PHILIC$),
   hydrophobic monomer ($\PHOBIC$), or vacuum ($\VACUUM$)\\
  \hline
   Molecular orientation $\MOri_\VECTOR{i}$
   & $\MOri_\VECTOR{i} \in L$ for polar molecules,
   $\MOri_\VECTOR{i} = \mathbf{null}$ otherwise\\
  \hline
   Translational kinetic energy $(\TKE_{\VECTOR{i}, l})_{l \in L}$ &
   $\TKE_{\VECTOR{i}, l} \in \{0,1\}$ for molecules,
   while non-zero values in the opposing directions are
   forbidden\\
  \hline
  Rotational kinetic energy $\RKE_\VECTOR{i}$ 
   & $\{-1, 0, +1\}$ (polar: $\{\pm 1\}$, non-polar: $0$)\\
  \hline
   Molecular bonds $(\MBond_{\VECTOR{i}, l})_{l \in L}$
   & Up to two bonds for hydrophilic or hydrophobic monomers
   \\
  \hline
  Heat particles $(\Heat_{\VECTOR{i}, l})_{l \in L}$ & $\Heat_{\VECTOR{i}, l} \in \{0,\ldots, \HeatMax\}$ for each $l \in L$\\
  \hline
  Preferential direction $\PDir_\VECTOR{i}$ 
   & $\PDir_\VECTOR{i} \in L$\\
  \hline
 \end{tabular}
}
\end{table}

For each cell $\VECTOR{i} \in \Z^2$, \NEWTERM{molecular type}
$\MType_\VECTOR{i}$ takes one of three types of
molecules---\STRESS{water} ($\WATER$), \STRESS{hydrophilic monomer}
($\PHILIC$), \STRESS{hydrophobic monomer} ($\PHOBIC$), or
\STRESS{vacuum} (empty; $\VACUUM$).
For example, one can consider the hydrophilic monomer to be acetic acid and
the hydrophobic monomer to be methane.
A site can contain at most one molecule; this constraint corresponds
to \NEWTERM{excluded volume}.
\begin{figure}[th]
\begin{center}
 \includegraphics[width=0.9\textwidth]{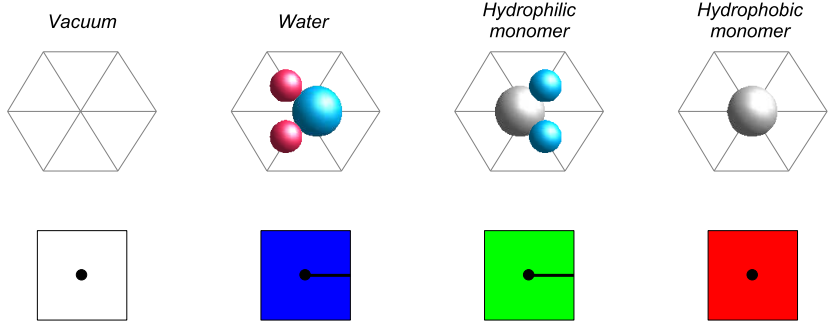}
 \caption{Molecular types. Upper row: schematic illustrations. Lower row:
 representations in visualization of the simulation results; bars
 indicate the orientations of polar molecules.}
 \label{fig:molecular_types}
\end{center}
\end{figure}

Water and hydrophilic monomers are polar molecules; therefore, they have
\NEWTERM{molecule orientation} $\MOri_\VECTOR{i} \in L$. (For
hydrophobic monomers and vacuum, $\MOri_\VECTOR{i} = \mathrm{null}$.)
We define that, for water in orientation $\MOri$, the same direction
represents negative polarization (corresponding to one oxygen) and
$\DirRotation^{\pm 2}(\MOri)$ represents positive polarization
(corresponding to two hydrogens), and for a hydrophilic monomer in
orientation $\MOri$, $\DirRotation^{\pm 1}(\MOri)$ represents negative
polarization (corresponding to O or OH) (See
Fig.~\ref{fig:molecular_types}).
Molecular orientation affects the strength of \STRESS{potential energy}
induced by several kinds of molecular interaction (see
section~\ref{subsec:potential}).

\begin{figure}[th]
 \begin{center}
 \includegraphics[width=0.9\textwidth]{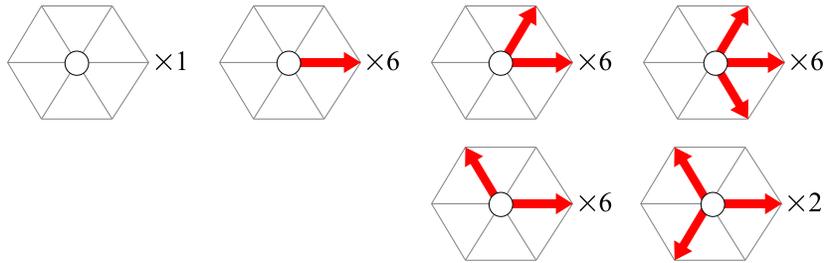}
  \caption{Possible states of translational kinetic energy for a molecule.}
  \label{fig:translational_kinetic_energies}
 \end{center}
\end{figure}

The sites occupied by molecules have \NEWTERM{translational kinetic
energy} (TKE) $\TKE_{\VECTOR{i}, l} \in \{0,1\}$ in every principal
direction $l \in L$, although non-zero energy values in opposite
directions on the same line are forbidden ($\TKE_{\VECTOR{i}, l} +
\TKE_{\VECTOR{i}, -l} \leq 1$).  Hence, there are $27$ possible TKE
states for a molecule (Fig.~\ref{fig:translational_kinetic_energies}).

Molecules can have \NEWTERM{rotational kinetic energy} (RKE)
$\RKE_\VECTOR{i}$, which allows the rotation of the polar molecules
to be reversible (see section~\ref{subsec:molecular_rotation}).
For proper update by the rotation rule given here, we confine the value
of $\RKE_\VECTOR{i}$ to $\{+1, -1\}$ for polar molecules, and to zero for
non-polar molecules or vacuum.
(For an alternative setting, see
Appendix~\ref{appsec:alternative_rotation}.)

\begin{figure}[th]
  \begin{center}
  \includegraphics[width=0.25\textwidth]{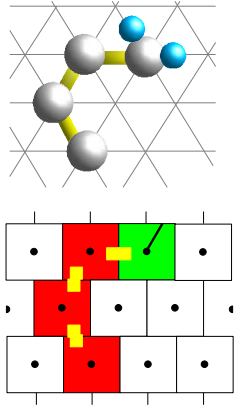}
   \caption{Schematic representation of a polymer.}
   \label{fig:illustration_polymer}
 \end{center}
\end{figure}

Hydrophilic and Hydrophobic monomers can have \NEWTERM{molecular bonds}
with neighboring monomers.
We define that 
\begin{equation}
 \label{eq:def_molecular_bonds}
  \MBond_{\VECTOR{i}, l} = 
  \left\{
   \begin{array}{cl}
    1 & \mbox{ if two monomers at $\VECTOR{i}$ and $l(\VECTOR{i})$ are bonded}, \\
    0 & \mbox{ otherwise}.
   \end{array}
  \right.  
\end{equation}
(Thus, $\MBond_{\VECTOR{i}, l} = \MBond_{l(\VECTOR{i}), -l}$.)
\NEWTERM{Polymers} can be composed as a group of monomers linked by the
bonds, as shown in Fig.~\ref{fig:illustration_polymer}.
In the current study, we suppose that for each monomer to have the bonds in at
most two directions, $\sum_{l \in L} \MBond_{\VECTOR{i}, l} \leq 2$;
thus, the polymers are one-dimensional.  One can consider the polymers to
be fatty acids.

For the above layers, which are related to molecules,
%
we also overlay the \NEWTERM{heat particles} layer on each cell.
The heat particle variable $\Heat_{\VECTOR{i}, l}$ can take values of $\{0,
\ldots, \HeatMax\}$ independently for every direction $l \in L$.

Finally, we append the \NEWTERM{preferential direction}
$\PDir_\VECTOR{i} \in L$ for each cell $\VECTOR{i}$.
In the transition rule given below, the preferential direction works as
a ``fluctuation'' to break irreversibility-inducing symmetry.
The \NEWTERM{parity} of the preferential direction, defined by 
\begin{equation}
 \label{eq:def_parity_pd}
 \PARITY(\PDir_\VECTOR{i}) = 
 \left\{
 \begin{array}{cl}
  +1 & \mathrm{if}\ \PDir_\VECTOR{i} \in \{+1, +3, -2\},\\
  -1 & \mathrm{if}\ \PDir_\VECTOR{i} \in \{+2, -1, -3\},\\
 \end{array}
 \right.
\end{equation}
is also utilized in the transition rule.

Molecular type, orientation, TKE and, molecular bonds (or variables
equivalent to them) are included in the original
LMA\cite{Mayer1998,Mayer2000}. On the other hand, RKE, heat particles,
and preferential direction are introduced in this model to implement
reversibility in a physically appropriate manner.

\subsection{Potential energy}\label{subsec:potential}

Every molecule interacts with its nearest neighboring
molecules;\footnote{Although wider range of interaction can also be
modeled, it requires larger number of \STRESS{site groups} and more
complicated update schemes (see section~\ref{subsec:transition_rule}).}
therefore, it has \NEWTERM{potential energies} for each of the six
principal directions. 
%
%
In calculating potential energy, we consider only pairwise interactions,
and let $V^{\VECTOR{i},l(\VECTOR{i})}$ denote potential energy arising
from the interaction between molecules at $\VECTOR{i}$ and
$l(\VECTOR{i})$.
The molecular interaction is divided into three classes:\footnote{We
omit cooperativity effects because of their minor influence on the
simulation results.}
\begin{itemize}
 \item Electrostatic interactions between permanent multipoles, 
       which take place when the polarized directions of the two polar
       molecules face each other.  
       %
       Let $V_{\mathrm{perm-perm}}$ represent the potential energy
       contribution from this class of interactions.

 \item Induction-based interactions between a permanent multipole and
       an induced multipole, 
       which take place when a polarized direction of one molecule faces
       an originally non-polarized direction of another.
       Let $V_{\mathrm{perm-ind}}$ represent the potential energy
       contribution from this class of interactions.

 \item London dispersion interactions between induced multipoles, which
       take place when the surfaces of two non-polar molecules face each
       other.
       Let $V_{\mathrm{ind-ind}}$ represent the potential energy
       contribution from this class of interactions.

\end{itemize}
Then, the total potential energy in the system is calculated as 
\begin{equation}
 \label{eq:def_total_potential}
 V_{\mathrm{total}}
 = \frac{1}{2} \sum_{\VECTOR{i}} \sum_{l \in L}  V^{\VECTOR{i},l(\VECTOR{i})}
 = \frac{1}{2} \sum_{\VECTOR{i}} \sum_{l \in L} 
 \left(
 V_{\mathrm{perm-perm}}^{\VECTOR{i},l(\VECTOR{i})}
 + V_{\mathrm{perm-ind}}^{\VECTOR{i},l(\VECTOR{i})}
 + V_{\mathrm{ind-ind}}^{\VECTOR{i},l(\VECTOR{i})}
 \right).
\end{equation}

For the full specification of the potential terms in our model, the
integer parameters listed in Table~\ref{tab:parameters_for_potential}
must be given.
\begin{table}[hbtp]
  \caption{Parameters of potential energy.
  Potentials of other neighboring directions of 
  molecular pairs are set to $0$.}
{
  \label{tab:parameters_for_potential}
  \begin{tabular}{|c|l|p{24em}|}
   \hline
   Class & Potential & Applied cases\\
   \hline
   & $V_{\mathrm{WH-WH}}$ & Where positively polarized directions
   (Hs)
   of two water molecules ($\WATER$s) face each other
   \\
   & $V_{\mathrm{WO-WO}}$ & Where negatively polarized directions
   (Os)
   of two $\WATER$s face each other
   \\
   & $V_{\mathrm{WH-WO}}$ & Where 
   an H and 
   an O of two $\WATER$s face each other
   \\
   \raisebox{1.em}{$V_{\mathrm{perm-perm}}$}
   & $V_{\mathrm{WH-IP}}$ & Where an H of a $\WATER$ faces 
   a negatively polarized direction (O or OH)
   of a hydrophilic monomer ($\PHILIC$)
   \\
   & $V_{\mathrm{WO-IP}}$ & Where an O of a $\WATER$ faces
   a negatively polarized direction 
   of an $\PHILIC$\\
   & $V_{\mathrm{IP-IP}}$ & Where two negatively polarized directions 
   of two $\PHILIC$s face each other
   \\
   \hline
   & $V_{\mathrm{WH-WN}}$ & Where an H and a non-polarized direction
   of two $\WATER$s face each other\\
   & $V_{\mathrm{WH-IN}}$ & Where an H of a $\WATER$ faces a
   non-polarized direction of an $\PHILIC$\\
   & $V_{\mathrm{WH-O}}$ & Where an H of a $\WATER$ faces any one of directions
   of a hydrophobic monomer ($\PHOBIC$)\\
   & $V_{\mathrm{WO-WN}}$ & Where an O and a non-polarized direction
   of two $\WATER$s face each other\\
   $V_{\mathrm{perm-ind}}$
   & $V_{\mathrm{WO-IN}}$ & Where an O of a $\WATER$ faces a
   non-polarized direction of an $\PHILIC$\\
   & $V_{\mathrm{WO-O}}$ & Where an O of a $\WATER$ faces any one of directions
   of an $\PHOBIC$\\
   & $V_{\mathrm{IP-WN}}$ & Where a negatively polarized direction of an
   $\PHILIC$ faces a non-polarized direction of a $\WATER$\\
   & $V_{\mathrm{IP-IN}}$ & Where a negatively polarized direction and
   a non-polarized direction of two $\PHILIC$s face each other\\
   & $V_{\mathrm{IP-O}}$ & Where a negatively polarized direction of an
   $\PHILIC$ faces any one of directions of $\PHOBIC$\\
   \hline
   $V_{\mathrm{ind-ind}}$
   & $V_{\mathrm{O-O}}$ & Where any directions of two $\PHOBIC$s face
   each other\\
   \hline
  \end{tabular}
}
\end{table}

\subsection{Transition rule}\label{subsec:transition_rule}

In the original LMA model, each unit-time update consists of the
following sub-steps\cite{Mayer1998}:
\begin{enumerate}
 \item propagation of the molecular type and redistribution of kinetic
       energies,

 \item construction of type-specific force fields,

 \item calculation of potential energies,

 \item calculation of the most proper move direction,

 \item readjustment of bonds in polymers according to the move
       direction, and

 \item movement of the molecule and clearing of the old lattice position.

\end{enumerate}

Although stated otherwise in Ref.~\cite{Mayer2000}, many of these
sub-steps are irreversible in actuality, involving erasure and
duplication of information. To realize reversibility, therefore, we
reconstruct the sub-steps and create new ones, utilizing the techniques
introduced in section~\ref{subsec:reversible_maps}.


\subsubsection{Molecular translation, collision, and excluded volume}\label{subsec:translation_collision_excluded_volume}

In the LMA, for each molecule, the most proper move direction
is calculated based on its TKEs and potentials, and the molecule moves
to the direction if the movement satisfies the constraints of excluded
volume and molecular bond maintenance.
This rule causes situations whose pre-images are not unique
(\textit{e.g.}, a molecule at a site might have come from one of the
neighboring sites according to the most proper move direction, or might
have been at the same site a unit time ago because of the constraints), and
thus it is irreversible.

\begin{figure}[th]
  \begin{center}
  \includegraphics[width=0.4\textwidth]{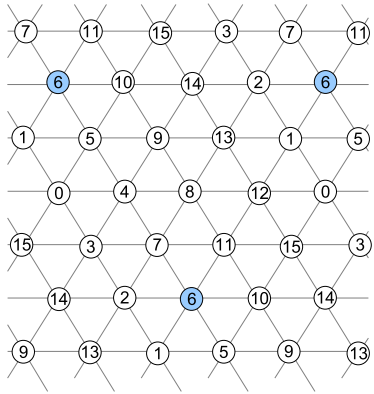}
   \caption{Site groups for interleaved interaction. Sites in the
   group indexed ``6'' are shaded.}
   \label{fig:TriangularLatticeInterleaved16}
 \end{center}
\end{figure}

To satisfy the constraints of excluded volume and molecular bond
maintenance, and to realize reversibility at the same time, we introduce
\NEWTERM{site groups}.  Sites in each group should be scattered
uniformly and sparsely enough (to prevent interference of the pairwise
interactions defined below, the sites in each group should be separated by
at least four times the unit distance).
We determine the group to which site $\VECTOR{i}$ belongs at time $t$ 
by the following map 
\begin{equation}
 \label{eq:def_site_group16}
  g(\VECTOR{i}, t) =
  \left\{4(i_x \bmod 4) + (i_y \bmod 4) + t\right\} \bmod 16,
\end{equation}
and let $G=\{0,1,\ldots, 15\}$ denote the range of $g$.
Here $(i_x, i_y)$ are the coordinates of site $\VECTOR{i}$ given by the
axis in Fig.~\ref{fig:triangular_lattice}(c), and each site is assigned to
a group, as shown in Fig.~\ref{fig:TriangularLatticeInterleaved16}.

Using the site groups and the preferential directions, 
molecular translation and collision are performed in an interleaved manner
using the scheme shown in Fig.\ref{fig:translation_collision_time_scheme}.
\begin{figure}[th]
 \begin{breakbox}
 \begin{tabbing}
  \textbf{begin}\\
 \textbf{for} $g$ in site groups $G$ \textbf{do}\\
  \hspace{4.0ex}\= 
    \textbf{for} $k$ in $(0, 1, \ldots, 5)$ \textbf{do}\\
  \>\hspace{4.0ex}\= 
  \textbf{for} every $\VECTOR{i}$ in a group $g$ \textbf{do}\\
  \>\>\hspace{4.0ex}\=
    $\VECTOR{j} := \PDir_\VECTOR{i}(\VECTOR{i})$; \\
  \>\>\>
    \textbf{if} $\DirRotation^{k}(\PDir_\VECTOR{j}) = \PDir_\VECTOR{i}$ \textbf{then}\\
  \>\>\>\hspace{4.0ex}\=
  $\PermMTC(\omega_\VECTOR{i},\omega_\VECTOR{j})$;
  \hspace{2ex}//pairwise interaction between $\VECTOR{i}$ and
  $\VECTOR{j}$.\\
  \textbf{end}
 \end{tabbing}
 \end{breakbox}
 \caption{Interleaved paired site interaction scheme. Note
 Eq.~(\ref{eq:def_dir_rotation_operator}) for the definition of
 $\DirRotation$.}
 \label{fig:translation_collision_time_scheme}
\end{figure}

For the interaction between paired neighboring sites $\VECTOR{i}$ and
$\VECTOR{j} = \PDir_\VECTOR{i}(\VECTOR{i})$ in
Fig.~\ref{fig:translation_collision_time_scheme}, we define a composite
conditional permutation $\PermMTC$, which represents molecular
translation and collision.
%

When both sites are vacuum, no interaction takes place:
\begin{equation}
\label{eq:pair_interaction_null}
 \MType_\VECTOR{i} = \MType_\VECTOR{j} = \VACUUM \Rightarrow 
 \PermMTC(\omega_\VECTOR{i}, \omega_\VECTOR{j})
 = (\omega_\VECTOR{i}, \omega_\VECTOR{j})  \mbox{ (identity).}
\end{equation}

When $\VECTOR{i}$ is occupied by a molecule and $\VECTOR{j}$ is
vacuum,
the molecule at $\VECTOR{i}$ moves to $\VECTOR{j}$ if the TKE in this
direction $\PDir_\VECTOR{i}$ is positive,
or the molecule's TKE in this direction is inverted if the TKE in the
opposite direction is positive, 
or no change takes place if the molecule's TKEs are zero in both
directions:
\begin{equation}
\label{eq:pair_interaction_translation}
 \begin{array}{l}
  \MType_\VECTOR{i} \neq \VACUUM, \ 
   \MType_\VECTOR{j} = \VACUUM \ \Rightarrow
   \\
  \left\{
   \begin{array}{rcl}
     \PermMTC\left(
     \left(
     \begin{array}{c}
      \MType_\VECTOR{i}\\
     \MOri_\VECTOR{i}\\
     (\TKE_{\VECTOR{i}, l})_l\\
     \RKE_\VECTOR{i}\\
     (\MBond_{\VECTOR{i}, l})_l \\
     \end{array}
     \right)
     ,
     \left(
     \begin{array}{c}
      \MType_\VECTOR{j}\\
     \MOri_\VECTOR{j}\\
     (\TKE_{\VECTOR{j}, l})_l\\
     \RKE_\VECTOR{j}\\
     (\MBond_{\VECTOR{j}, l})_l \\
     \end{array}
     \right)
     \right)
     &=&
     \left(
     \left(
     \begin{array}{c}
      \MType_\VECTOR{j}\\
     \MOri_\VECTOR{j}\\
     (\TKE_{\VECTOR{j}, l})_l\\
     \RKE_\VECTOR{j}\\
     (\MBond_{\VECTOR{j}, l})_l \\
     \end{array}
     \right)
     , 
     \left(
     \begin{array}{c}
      \MType_\VECTOR{i}\\
     \MOri_\VECTOR{i}\\
     (\TKE_{\VECTOR{i}, l})_l\\
     \RKE_\VECTOR{i}\\
     (\MBond_{\VECTOR{i}, l})_l \\
     \end{array}
     \right)
     \right)\\
    &&\ \mbox { if } 
     (\TKE_{\VECTOR{i}, \PDir_\VECTOR{i}},
     \TKE_{\VECTOR{i}, -\PDir_\VECTOR{i}})= (1,0),
     \\
%
     \PermMTC(\TKE_{\VECTOR{i}, \PDir_\VECTOR{i}}, 
     \TKE_{\VECTOR{i}, -\PDir_\VECTOR{i}}) &=&
     (\TKE_{\VECTOR{i},-\PDir_\VECTOR{i}}, 
     \TKE_{\VECTOR{i}, \PDir_\VECTOR{i}})
     \\
     &&\ \mbox { if } 
     (\TKE_{\VECTOR{i}, \PDir_\VECTOR{i}},
     \TKE_{\VECTOR{i}, -\PDir_\VECTOR{i}}) = (0,1),
     \\
%
     \PermMTC(\omega_\VECTOR{i}, \omega_\VECTOR{j})
     &=& (\omega_\VECTOR{i}, \omega_\VECTOR{j})\\
     &&\ \mbox { if } 
     (\TKE_{\VECTOR{i}, \PDir_\VECTOR{i}},
     \TKE_{\VECTOR{i}, -\PDir_\VECTOR{i}}) = (0,0)
     .\\
   \end{array}  
\right.
 \end{array}
\end{equation}
(Here, internal layers that are unaffected by $\PermMTC$ are omitted.)
Conversely, when $\VECTOR{i}$ is vacuum and $\VECTOR{j}$ is occupied by
a molecule, the molecule's move direction $\PDir_\VECTOR{i}$ is replaced
by $-\PDir_\VECTOR{i}$ in the above equation.
Note that 
%
if the preferential directions are uniform over all of the sites, a free
molecule without molecular interaction always maintains the directions of
its TKEs after one full unit-time update, because whenever an inversion
of TKE takes place, it is canceled by another inversion in the update
scheme of Fig.~\ref{fig:translation_collision_time_scheme}.
This does not hold, however, if the preferential directions are not uniform.
This issue will be addressed again in section~\ref{subsec:conservation_laws}.

When both $\VECTOR{i}$ and $\VECTOR{j}$ are occupied by molecules,
the two molecules exchange TKEs as in an elastic collision
\begin{equation}
\label{eq:pair_interaction_collision}
 \begin{array}{l}
  \mbox{$\MType_\VECTOR{i} \neq \VACUUM \wedge \MType_\VECTOR{j} \neq \VACUUM$}\\
  \qquad \Rightarrow 
   \PermMTC(
   \TKE_{\VECTOR{i}, \pm\PDir_\VECTOR{i}}, 
   \TKE_{\VECTOR{j}, \pm\PDir_\VECTOR{i}}
   ) = (
   \TKE_{\VECTOR{j}, \pm\PDir_\VECTOR{i}}, 
   \TKE_{\VECTOR{i}, \pm\PDir_\VECTOR{i}}
   ). \\
 \end{array}
\end{equation}

\subsubsection{Maintenance of molecular bonds}\label{subsec:molecular_bonds_sustainability}

As mentioned earlier, polymers are composed as chains of monomers linked
by molecular bonds (Fig.~\ref{fig:illustration_polymer}). 
Integrated and coherent motion of such a multi-site structure
(``solid body'') is difficult to model using CA.
A possible approach is to express the structure's motion states by its
deforming shape\cite{Chopard1990,Marconi2003}. Although this method can
replicate many aspects of Hamiltonian mechanics as well as the
structure's integrity, it makes it difficult to formalize proper interaction
between such a structure and the single-site particles
(molecules) whose motion states are expressed as their internal states.
In the RLMA, the maintenance of bonds is ensured by using the bond
information as another conditional signal for the molecular translational
permutation (\ref{eq:pair_interaction_translation}).

\begin{figure}[th]
 \begin{center}
  \includegraphics[width=0.9\textwidth]{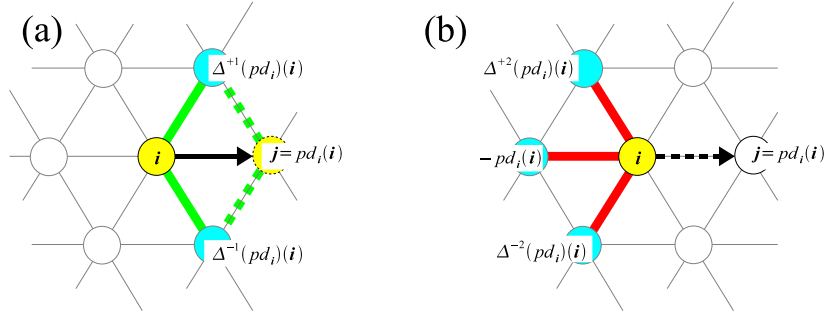}
  \caption{Translational movement and molecular bonds. If a molecule at
  $\VECTOR{i}$ is moving to $\VECTOR{j} =
  \PDir_{\VECTOR{i}}(\VECTOR{i})$, (a) its bonds in directions
  $\DirRotation^{\pm 1}(\PDir_{\VECTOR{i}})$ are maintained with their
  directions readjusted to $\DirRotation^{\pm 2}(\PDir_{\VECTOR{i}})$,
  but (b) bonds in directions $\DirRotation^{\pm 2}(\PDir_{\VECTOR{i}})$
  and $-\PDir_{\VECTOR{i}}$ would not be maintained.}
  \label{fig:MolecularBondsMaintenance}
 \end{center}
\end{figure}

First, it is checked if the molecular bonds (provided they exist) are
maintained when the molecule at $\VECTOR{i}$ moves to the vacuum site
$\VECTOR{j}=\PDir_{\VECTOR{i}}(\VECTOR{i})$.
Bonds with molecules at $\DirRotation^{\pm
1}(\PDir_{\VECTOR{i}})(\VECTOR{i})$ are not destroyed by the movement
because the bonded molecules remain neighbors (the bond directions after
movement become $\DirRotation^{\pm 2}(\PDir_{\VECTOR{i}})$;
Fig.~\ref{fig:MolecularBondsMaintenance}(a)).  If the molecule has bonds
in other directions, these bonds are destroyed by the movement
(Fig.~\ref{fig:MolecularBondsMaintenance}(b)).  Therefore, we append
another condition to the permutation
(\ref{eq:pair_interaction_translation}):
\begin{equation}
\label{eq:pair_interaction_bond_maintenance_condition}
 \begin{array}{l}
  \MType_\VECTOR{i} \neq \VACUUM, \ \MType_\VECTOR{j} = \VACUUM
   \Rightarrow \\
   \left\{
   \begin{array}{cl}
    \mbox{Apply (\ref{eq:pair_interaction_translation}) followed by bond readjustment } \BondReadjustment& 
     \mbox{ if }
    \MBond_{\VECTOR{i}, \DirRotation^{\pm 2}(\PDir_{\VECTOR{i}})} = \MBond_{\VECTOR{i}, -\PDir_{\VECTOR{i}}} = 0,\\
     \PermMTC(\omega_\VECTOR{i}, \omega_\VECTOR{j})
     = (\omega_\VECTOR{i}, \omega_\VECTOR{j}) 
     & \mbox{ otherwise}.
    \\
   \end{array}  
\right.
 \end{array}
\end{equation}
The readjustment of bonds $\BondReadjustment$ takes place not only at
the moved molecule's new position $\VECTOR{j}$ but also at
$\DirRotation^{\pm 1}(\PDir_{\VECTOR{i}})(\VECTOR{i})$ if the bonds
exist:
\begin{equation}
\label{eq:bonds_readjustment}
    \BondReadjustment\left(
	     \begin{array}{c}
	      \MBond_{\VECTOR{j}, \DirRotation^{+1}(\PDir_{\VECTOR{i}})},
	       \MBond_{\VECTOR{j}, \DirRotation^{+2}(\PDir_{\VECTOR{i}})}\\
	      \MBond_{\DirRotation^{+1}(\PDir_{\VECTOR{i}})(\VECTOR{i}), \DirRotation^{-2}(\PDir_{\VECTOR{i}})}, 
	       \MBond_{\DirRotation^{+1}(\PDir_{\VECTOR{i}})(\VECTOR{i}), \DirRotation^{-1}(\PDir_{\VECTOR{i}})}\\
	      \MBond_{\VECTOR{j}, \DirRotation^{-1}(\PDir_{\VECTOR{i}})},
	       \MBond_{\VECTOR{j}, \DirRotation^{-2}(\PDir_{\VECTOR{i}})}\\
	      \MBond_{\DirRotation^{-1}(\PDir_{\VECTOR{i}})(\VECTOR{i}), \DirRotation^{+2}(\PDir_{\VECTOR{i}})}, 
	       \MBond_{\DirRotation^{-1}(\PDir_{\VECTOR{i}})(\VECTOR{i}), \DirRotation^{+1}(\PDir_{\VECTOR{i}})}\\
	     \end{array}
	    \right) 
    = 
    \left(
     \begin{array}{c}
      0,
       \MBond_{\VECTOR{j}, \DirRotation^{+1}(\PDir_{\VECTOR{i}})}\\
      0, 
       \MBond_{\DirRotation^{+1}(\PDir_{\VECTOR{i}})(\VECTOR{i}), \DirRotation^{-2}(\PDir_{\VECTOR{i}})}\\
      0,
       \MBond_{\VECTOR{j}, \DirRotation^{-1}(\PDir_{\VECTOR{i}})}\\
      0,
       \MBond_{\DirRotation^{-1}(\PDir_{\VECTOR{i}})(\VECTOR{i}), \DirRotation^{+2}(\PDir_{\VECTOR{i}})}\\
     \end{array}
    \right), \\
\end{equation}

A drawback of this rule is that it occasionally causes motion of
polymers that has less physical relevance. 
For example, when a polymer's constituent monomers are arranged on a
straight line and all of the monomers have positive TKE only 
on the line, the polymer cannot move even if all of the monomers' TKE
directions are identical.
However, if some of the monomers have positive TKEs in other directions,
this polymer can move on average to its most proper direction with
respect to TKE, while becoming deformed and keeping its integrity.

\subsubsection{Self-organization and reversibility}\label{subsec:self_organization_and_reversibility}

In irreversible models such as the LMA, the tendency of
self-organization from disordered high-entropy states to ordered
low-entropy structures is embedded in their information-losing
transition rules themselves.
To realize ``apparent'' self-organization of ordered structures by a
reversible rule without information loss, the rule needs additional
degrees of freedom that work as a \STRESS{heat bath} into which the
entropy generated in the organization process should be disposed of.
The deterministic Ising model\cite{Creutz1986} and the reversible
generalization\cite{D'Souza1999} of the diffusion limited aggregation
(DLA) model\cite{Witten1981} are examples of this approach.
We also adopt this approach, using the \STRESS{heat particle layer} as
the heat bath.


%

When both of the neighboring sites $\VECTOR{i}$ and $\VECTOR{j}$
have molecules, in advance of the collision
(\ref{eq:pair_interaction_collision}) by $\PermMTC$, 
we apply site-respective TKE--heat interaction $\PermMolHeat$ defined as
follows. 
For the molecule at site $\VECTOR{x}$ (i.e., either $\VECTOR{i}$ or
$\VECTOR{j}$),
when it has positive TKE in one direction $l$ out of $\{\pm
\PDir_{\VECTOR{i}} \}$ (i.e., along the line connecting the two
molecules), and when the heat particles satisfy $\Heat_{\VECTOR{x}, l} <
H_{\mathit{max}}$ and $\Heat_{\VECTOR{x}, -l} = 0$ at site $\VECTOR{x}$,
then the positive TKE is transformed into a heat particle in the same
direction $l$ (heat release):
\begin{equation}
 \label{eq:release_heat_before_pair_interaction}
 \begin{array}{l}
  \MType_\VECTOR{i} \neq \VACUUM, \ \MType_\VECTOR{j} \neq \VACUUM 
  \mbox{ then for } \VECTOR{x} \in \{\VECTOR{i}, \VECTOR{j}\}\\
  \left\{
   \begin{array}{cl}
    \PermMolHeat
    \left(
     \begin{array}{c}
      \TKE_{\VECTOR{x}, \pm\PDir_{\VECTOR{i}}}
       \\
      \Heat_{\VECTOR{x}, \pm\PDir_{\VECTOR{i}}}
     \end{array}    
    \right)
    = 
    \left(
     \begin{array}{c}
      \TKE_{\VECTOR{x}, \pm\PDir_{\VECTOR{i}}}-1
       \\
      \Heat_{\VECTOR{x}, \pm\PDir_{\VECTOR{i}}}+1
     \end{array}    
    \right)
    &
     \begin{array}{l}
     \mbox{ if }
    (\TKE_{\VECTOR{x}, \pm\PDir_{\VECTOR{i}}}, 
       \TKE_{\VECTOR{x}, \mp\PDir_{\VECTOR{i}}}) = (1,0)\\
     \quad\mbox{ and }
      \Heat_{\VECTOR{x}, \pm\PDir_{\VECTOR{i}}} < \HeatMax, 
      \ \Heat_{\VECTOR{x}, \mp\PDir_{\VECTOR{i}}} = 0,\\
     \end{array}\\
    \PermMolHeat(\omega_\VECTOR{x}) = \omega_\VECTOR{x} &\mbox{ otherwise}.
   \end{array}
  \right.
 \end{array}
\end{equation}
On the other hand, when the molecule at $\VECTOR{x}$ does not have a TKE
in either of $\{\pm \PDir_{\VECTOR{i}} \}$ and heat particles exist in
only one of the two directions, one heat particle is transformed into
the TKE in the same direction (heat absorption):
\begin{equation}
 \label{eq:absorb_heat_before_pair_interaction}
 \begin{array}{l}
  \MType_\VECTOR{i} \neq \VACUUM, \ \MType_\VECTOR{j} \neq \VACUUM 
  \mbox{ then for } \VECTOR{x} \in \{\VECTOR{i}, \VECTOR{j}\}\\
  \left\{
   \begin{array}{cl}
    \PermMolHeat
    \left(
     \begin{array}{c}
      \TKE_{\VECTOR{x}, \pm\PDir_{\VECTOR{i}}}
       \\
      \Heat_{\VECTOR{x}, \pm\PDir_{\VECTOR{i}}}
     \end{array}    
    \right)
    = 
    \left(
     \begin{array}{c}
      \TKE_{\VECTOR{x}, \pm\PDir_{\VECTOR{i}}}+1
       \\
      \Heat_{\VECTOR{x}, \pm\PDir_{\VECTOR{i}}}-1
     \end{array}    
    \right)
    &
     \begin{array}{l}
     \mbox{ if }
    (\TKE_{\VECTOR{x}, \PDir_{\VECTOR{i}}}, 
       \TKE_{\VECTOR{x}, -\PDir_{\VECTOR{i}}}) = (0,0)\\
     \quad\mbox{ and }
      \Heat_{\VECTOR{x}, \pm\PDir_{\VECTOR{i}}} > 0,
      \ \Heat_{\VECTOR{x}, \mp\PDir_{\VECTOR{i}}} = 0,\\
     \end{array}\\
    \PermMolHeat(\omega_\VECTOR{x}) = \omega_\VECTOR{x} &\mbox{ otherwise}.
   \end{array}
  \right.
 \end{array}
\end{equation}
Introduction of this TKE--heat interaction makes the interaction between
neighboring molecules non-elastic.


For molecular translation, potentials work as yet another control
signal for the permutation in $\PermMTC$ of
Eq.~(\ref{eq:pair_interaction_translation}), and the translation 
is executed only if the changes in potentials entailed by the movement
of molecule from $\VECTOR{i}$ to $\VECTOR{j}$, given that their
neighbors are fixed, can be compensated for by emission/absorption of
heat particles at $\VECTOR{i}$ and $\VECTOR{j}$:
\begin{equation}
 \label{eq:translation_condition_potential_change_compensation}
 \begin{array}{l}
  \MType_\VECTOR{i} \neq \VACUUM$, $\MType_\VECTOR{j} = \VACUUM \Rightarrow \\
   \left\{
   \begin{array}{cl}
    \mbox{Proceed to (\ref{eq:pair_interaction_bond_maintenance_condition})}
     & 
     \mbox{ if }
     0 \leq \Heat_{\VECTOR{i}, l} - \PotentialChangeByMolRemoval^{\VECTOR{i}, l(\VECTOR{i})} \leq \HeatMax 
     \mbox{ and } \\
    & \qquad
     0 \leq \Heat_{\VECTOR{j}, l} - \PotentialChangeByMolPutting^{\VECTOR{j}, l(\VECTOR{j})} \leq \HeatMax \mbox{ for } l \in L,
     \\
     \PermMTC(\omega_\VECTOR{i}, \omega_\VECTOR{j})
     = (\omega_\VECTOR{i}, \omega_\VECTOR{j}) 
     & \mbox{ otherwise}.
    \\
   \end{array}  
\right.
 \end{array}
\end{equation}
Here, $\PotentialChangeByMolRemoval^{\VECTOR{i}, l(\VECTOR{i})}$
represents the potential change in direction $l$ caused by removing the
molecule from its current site $\VECTOR{i}$, and
$\PotentialChangeByMolPutting^{\VECTOR{j}, l(\VECTOR{j})}$ represents a
potential change in direction $l$ caused by placing the molecule
(while maintaining its orientation) at vacuum site $\VECTOR{j}$.
%
Further, when the translation is actually induced by the permutation
(\ref{eq:pair_interaction_translation}), it is followed by the potential
change compensation $\PotentialChangeCompensationByHeats$:
\begin{equation}
 \label{eq:potential_change_compensation_translation}
 \PotentialChangeCompensationByHeats
  \left(
   (\Heat_{\VECTOR{i},l})_{l}, (\Heat_{\VECTOR{j},l})_{l}
  \right)
  = 
  \left(
   (\Heat_{\VECTOR{i},l} - \PotentialChangeByMolRemoval^{\VECTOR{i}, l(\VECTOR{i})})_{l}, 
   (\Heat_{\VECTOR{j},l} - \PotentialChangeByMolPutting^{\VECTOR{j}, l(\VECTOR{j})})_{l}
  \right).
\end{equation}

Thus, a molecule moving to a more stable site
($\sum_{l}\PotentialChangeByMolPutting^{\VECTOR{j}, l(\VECTOR{j})} < 0$)
releases heat particles in total, and it can be dissociated again from
its neighbors only when enough energy is supplied by the heat particle
layer.

\subsubsection{Rotation of polar molecules}\label{subsec:molecular_rotation}

Polar molecules such as water and hydrophilic monomers can take different
potential values depending on their orientations.
According to the LMA rule, the polar molecules are rotated
irreversibly into their more stable (lower-potential) orientations.
To maintain reversibility and at the same time to enable relaxation into
a more stable direction configuration, we utilize RKE and also the heat
particle layer. 
Similar to the paired site interaction in
Fig.~\ref{fig:translation_collision_time_scheme}, rotational update is
also performed in the interleaved scheme shown in
Fig.~\ref{fig:molecular_rotation_time_scheme}.  Here, we again use the
site groups of Eq.~(\ref{eq:def_site_group16}) (although for the
rotational permutation given below, interference can be prevented if the
sites in each group are separated by more than a unit distance).
\begin{figure}[th]
 \begin{breakbox}
 \begin{tabbing}
  \textbf{start}\\
 \textbf{for} $g$ in site groups $G$ \textbf{do}\\
  \hspace{4.0ex}\= 
  \textbf{for} every $\VECTOR{i}$ in a group $g$ \textbf{do}\\
  \>\hspace{4.0ex}\=
  \textbf{if} $\MType_\VECTOR{i} \in \{\WATER, \PHILIC\}$ \textbf{then} 
  \hspace{2ex} //polar molecule\\
  \>\>\hspace{4.0ex}\=  $\PermRot(\MOri_\VECTOR{i}, \RKE_\VECTOR{i}, (\Heat_{\VECTOR{i},l})_{l \in L})$;\hspace{2ex} //rotational update \\
\textbf{end}
 \end{tabbing}
 \end{breakbox}
 \caption{Interleaved update scheme for molecular orientation and RKE
 with heat interaction.}
 \label{fig:molecular_rotation_time_scheme}
\end{figure}

The rotational update $\PermRot$ is defined as follows:
the polar molecule at $\VECTOR{i}$ rotates according to the sign of RKE
and the orientation becomes
$\DirRotation^{\RKE_\VECTOR{i}}(\MOri_\VECTOR{i})$ if the change in
potentials caused by the rotation can be compensated for by
emission/absorption of heat particles at the site.
Otherwise, the molecule does not rotate and RKE is inverted:
\begin{equation}
 \label{eq:rotational_update}
 \PermRot\left(
	  \begin{array}{c}
	   \MOri_\VECTOR{i} \\
	   \RKE_\VECTOR{i} \\
	   (\Heat_{\VECTOR{i}, l})_l
	  \end{array}
	 \right)
 = 
 \left\{
 \begin{array}{l}
  \left(
   \begin{array}{c}
    \DirRotation^{\RKE_\VECTOR{i}}(\MOri_\VECTOR{i})\\
    \RKE_\VECTOR{i}\\
     (\Heat_{\VECTOR{i}, l}
      - \PotentialChangeByRotation{\MOri_\VECTOR{i}}{\RKE_\VECTOR{i}}^{\VECTOR{i}, l(\VECTOR{i})})_l
      \end{array}  
  \right)
  \\
  \qquad \mbox{ if }
   0 \leq
   \Heat_{\VECTOR{i}, l}
      - \PotentialChangeByRotation{\MOri_\VECTOR{i}}{\RKE_\VECTOR{i}}^{\VECTOR{i}, l(\VECTOR{i})}
      \leq \HeatMax
   \mbox{ for } {}^\forall l \in L, \\
  \left(
   \begin{array}{c}
    \MOri_\VECTOR{i}\\
    - \RKE_\VECTOR{i}\\
    (\Heat_{\VECTOR{i}, l})_l
   \end{array}
    \right) 
   \qquad \mbox{ otherwise}.\\
 \end{array}
 \right.
\end{equation}
Here 
\begin{equation}
 \label{eq:def_potential_change_by_rotation}
 \PotentialChangeByRotation{k}{n}^{\VECTOR{i}, l(\VECTOR{i})}
\end{equation}
represents the potential change in direction $l$ that occurs when the
orientation of the molecule at $\VECTOR{i}$ is changed from $k$ to 
$\DirRotation^{n}(k)$, with the neighboring molecules fixed.

In this rule, the RKEs work as a kind of ``second-order''
signal\cite{Toffoli1990}, which preserves the history of the molecules'
rotational states.
Therefore, the polar molecules cannot stop rotating i.e., $\RKE \neq 0$
(or given $\RKE \neq 0$ as the initial condition, they cannot change
their orientations forever).
For an alternative rotation rule that allows RKEs to take values on
$\Z$, including the stationary state $\RKE = 0$,
see Appendix \ref{appsec:alternative_rotation}.

\subsubsection{Transportation of heat particles}

For the heat particle layer to function as a heat bath, the released
heat particles should be effectively diffused
into open areas.  In our RLMA model, the diffusion of heat particles is
conducted by a rule similar to that for the Frisch--Hasslacher--Pomeau
lattice gas automata (FHP-LGA)\cite{Frisch1986}, that is, for every
unit-time update, a synchronous shift (translation) $\HeatShift$ is
performed in each direction
\begin{equation}
 \label{eq:heat_shift}
  \HeatShift: h_{\VECTOR{i},l} \mapsto  h_{-l(\VECTOR{i}),l}
  \quad \mbox{for}\quad l \in L
\end{equation}
followed by a local collision $\HeatCollision$ at each site:
\begin{equation}
 \label{eq:heat_collision}
  \HeatCollision: 
  \left\{
   \begin{array}{ccl}
    \left( (h_{\VECTOR{i},\DirRotation^{l}(k)}) \right)  = 
     (m, 0, 0, m, 0, 0)
     &\mapsto& 
     \left\{
      \begin{array}{cl}
    (0, m, 0, 0, m, 0) & \mbox{ if } \PARITY(\PDir_\VECTOR{i}) = +1, \\
    (0, 0, m, 0, 0, m) & \mbox{ if } \PARITY(\PDir_\VECTOR{i}) = -1, \\
      \end{array}
     \right.
     \\
    \left( (h_{\VECTOR{i},\DirRotation^{l}(k)}) \right)  = 
     (m, 0, m, 0, m, 0)
     &\mapsto& 
    (0, m, 0, m, 0, m) \\
    \left( (h_{\VECTOR{i},\DirRotation^{l}(k)}) \right)
     &\mapsto&
     \left( (h_{\VECTOR{i},\DirRotation^{l}(k)}) \right) \quad\mbox{ otherwise}
   \end{array}
   \right.
\end{equation}
with $0 < m \leq \HeatMax$ and $k \in \{1,2,3\}$.
Note that the collision $\HeatCollision$ is deterministic and utilizes
the parity of preferential direction at each site.

\subsection{Composition of unit-time update}\label{subsec:new_model_construction}

After integrating the sub-steps defined above, a unit-time update of the
RLMA can be performed through the following time sub-steps (variables in
parentheses are those affected by the particular sub-step):
\begin{enumerate}
 \item Transportation of heat particles ($(\Heat_{\VECTOR{i},l}), \PDir_\VECTOR{i}$)

       Heat particles are diffused by the FHP-LGA-like combination of
       the shift (\ref{eq:heat_shift}) and collision
       (\ref{eq:heat_collision}).

 \item Rotation of polar molecules ($\MType_\VECTOR{i}, \MOri_\VECTOR{i}, \RKE_\VECTOR{i}, (\Heat_{\VECTOR{i},l})$)

       Using the interleaved scheme of
       Fig.~\ref{fig:molecular_rotation_time_scheme}, molecular rotation
       is performed by the conditional permutation
       (\ref{eq:rotational_update}).

 \item Molecular translation and interaction ($\MType_\VECTOR{i}, \MOri_\VECTOR{i}, (\TKE_{\VECTOR{i},l}),
       (\MBond_{\VECTOR{i},l}), (\Heat_{\VECTOR{i},l}), \PDir_\VECTOR{i}$)

       Using the interleaved scheme of
       Fig.~\ref{fig:translation_collision_time_scheme}, 
       \begin{itemize}
	\item molecular translation is performed by the paired site
	      conditional permutation
	      (\ref{eq:pair_interaction_translation}) with the
	      conditions
	      (\ref{eq:pair_interaction_bond_maintenance_condition})
	      and
	      (\ref{eq:translation_condition_potential_change_compensation}),
	      while

	\item molecular interaction is performed by the conditional
	      permutation (\ref{eq:pair_interaction_collision}) with 
	      the heat release
	      (\ref{eq:release_heat_before_pair_interaction}) and absorption
	      (\ref{eq:absorb_heat_before_pair_interaction}).  
       \end{itemize}

 \item Update of preferential direction ($\PDir_\VECTOR{i}$)

       To ensure unbiasedness for the principal directions, the
       preferential direction should be updated according to time.  
       %
       %
       We use the simple uniform rotation $\PDir_\VECTOR{i} \mapsto
       \DirRotation^{+1}(\PDir_\VECTOR{i})$, although synchronous shifts
       and deterministic, invertible pseudorandom number generators can
       also be combined.
\end{enumerate}
These sub-steps are independent; therefore, the order can be changed.
Each sub-time step is reversible; therefore, the inverse update is
achieved by performing this construction in reverse.

\subsection{Conservation laws}
\label{subsec:conservation_laws}

From the above definitions, the transition rule of RLMA conserves
\STRESS{mass} (number of molecules) and the \NEWTERM{total energy}
that is given as a sum of TKEs, RKEs, potential energies, and heat
particles over the sites:
\begin{equation}
 \label{eq:energy_conservation_law}
 E_{\mathrm{total}}
 = \sum_{\VECTOR{i}}\sum_{l \in L} \TKE_{\VECTOR{i},l}
 + \sum_{\VECTOR{i}} |\RKE_\VECTOR{i}|
 + V_{\mathrm{total}}
 + \sum_{\VECTOR{i}}\sum_{l \in L} \Heat_{\VECTOR{i},l}.
\end{equation}
These conservation laws enable precise application and validation of
methods and theorems in statistical mechanics, both equilibrium
(microcanonical, canonical, and grand-canonical) and nonequilibrium
(\textit{e.g.}, relaxation, Fourier's law of heat conduction,
Green--Kubo relations), as is done for simpler (and in many cases more
abstract) CA
models\cite{Creutz1986,D'Souza1999,Takesue1990,Takesue1990a,Takesue1997,Niwa1997,Saito1999,Stauffer2000}.

On the other hand, 
conservation of momenta does not hold 
because of the TKE inversion in the translational permutation
(\ref{eq:pair_interaction_translation}), as mentioned in
section~\ref{subsec:translation_collision_excluded_volume}.  (Angular
momenta are not conserved either, because the rotational permutation
(\ref{eq:rotational_update}) also contains uncompensated inversion of
RKE.)
From the macroscopic viewpoint, this non-conservation of momenta seems to
work positively 
to enhance the model's ergodicity, instead of the effect
of chaos dynamics, which discrete CA models lack.

\section{Simulation}\label{sec:simulation}

In this section, we demonstrate by simulation that 
the RLMA can reproduce the original LMA's self-organization
results\cite{Mayer1997} as special cases.
We also show that the RLMA reproduces results which are qualitatively
consistent with the traditional Larson-type models.
More extensive results and their statistical treatment will be given in
a future work.

In the following simulations, the parameters are set as follows:
$\HeatMax = 8$, 
$V_\mathrm{WH-WH} = V_\mathrm{WO-WO}= V_\mathrm{WO-IP} = V_\mathrm{IP-IP} = +4$, 
$V_\mathrm{WH-WO} = V_\mathrm{WH-IP} = -4$,
$V_\mathrm{WH-WN} = V_\mathrm{WH-IN}= V_\mathrm{WH-O} = 
V_\mathrm{WO-WN} = V_\mathrm{WO-IN}= V_\mathrm{WO-O} = 
V_\mathrm{IP-WN} = V_\mathrm{IP-IN}= V_\mathrm{IP-O} = 
V_\mathrm{O-O} = -1$. 
The simulations in section~\ref{subsec:water_phobic_clustering} and
\ref{subsec:polymer_dynamics} adopted lattice space consisting of
$N= 24\times24$ cells, while the simulations in
section~\ref{subsec:ternary_mixtures} used lattice space of $N= 100 \times
100$ cells.
Periodic boundary condition is applied in all the simulations, 
so the systems are isolated.

\subsection{Hydrophobic monomers in a polar environment}\label{subsec:water_phobic_clustering}

\begin{figure}[th]
 \begin{center}
 \includegraphics[width=1.0\textwidth]{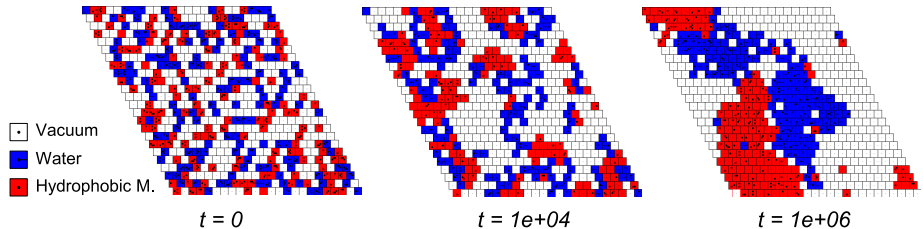}
  \caption{Snapshots of the molecular layer in a simulation of
  a water--hydrophobic monomer system. The molecules cluster
  preferentially with those of the same type, and phase separation 
  occurs.}
  \label{fig:molecular_aggregation_water_phobic}
 \end{center}
\end{figure}

Fig.~\ref{fig:molecular_aggregation_water_phobic} shows 
snapshots of the molecular layer in a simulation of 
a mixture of water and hydrophobic monomers
(25\% water, 25\% hydrophobic monomer, 50\% vacuum).
%
%
Starting from a homogeneously mixed initial configuration with no heat
particles (Fig.~\ref{fig:molecular_aggregation_water_phobic}(a)), 
clustering and phase separation gradually take place
(Fig.~\ref{fig:molecular_aggregation_water_phobic}(b), (c)), accompanied
by emission of heat particles (not shown in the figures).

Note also that phase separation takes place in spite of the setup that
the induction-based forces $v_\mathrm{WO-H}$ and $v_\mathrm{WO-O}$
between a hydrophobic monomer and a water molecule are set equal to the
dispersion interaction force $v_\mathrm{OO}$ between two hydrophobic
monomers, and they are much weaker than the water--water binding
$v_\mathrm{WH-WO}$, as is the case in Ref.~\cite{Mayer1997} (see also
section~\ref{subsec:LMAandLarson}).
Since the system is isolated, this self-organization process is
\STRESS{entropy-driven}.

\begin{figure}[th]
 \begin{center}
 \includegraphics[width=0.45\textwidth]{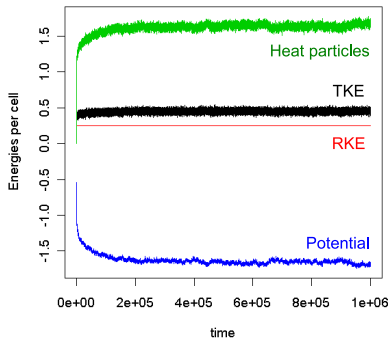}
 \caption{Time evolution of mean values of TKE, RKE, potential energy,
 and heat particles per cell in a simulation of a water--hydrophobic
 monomer system.}
  \label{fig:energies_transition_water_phobic}
 \end{center}
\end{figure}

Fig.~\ref{fig:energies_transition_water_phobic} shows the time evolution of
mean energies per cell---TKE (sum for all of the principal directions
$\langle \sum_{l \in L} \TKE_{\VECTOR{i},l} \rangle_{\VECTOR{i}}$), RKE
(absolute value $\langle |\RKE_\VECTOR{i}| \rangle_{\VECTOR{i}}$),
potential energy (sum for all of the principal directions
$\frac{1}{2}\langle \sum_{l \in L} V^{\VECTOR{i},l}
\rangle_{\VECTOR{i}}$), heat particles (sum for all of the principal
directions $\langle \sum_{l \in L} \Heat_{\VECTOR{i},l}
\rangle_{\VECTOR{i}}$)---in the simulation run of
Fig.~\ref{fig:molecular_aggregation_water_phobic}.
In this relaxation process, as the molecules are organized into a more
stable configuration, the energy released from the molecular layer is
transferred into the heat particle layer, with the total energy
conserved (mean total energy per cell, $e_{\mathrm{total}} =
E_{\mathrm{total}}/N$, is $0.68$).
It is observed that a large part of the energy transfer takes place in
the first few thousand steps.  This corresponds to quick dissolution of
high-potential, unstable partial configurations.

\begin{figure}[th]
 \begin{center}
 \includegraphics[width=0.45\textwidth]{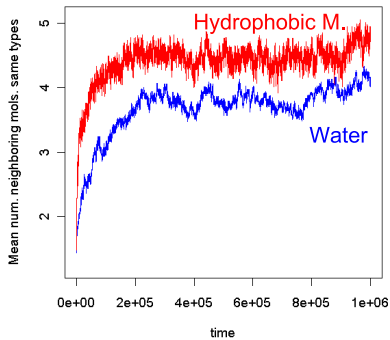}
  \caption{Time evolution of mean numbers of neighboring molecules of
  the same types (water and hydrophobic monomer) in the simulation of
  a water--hydrophobic monomer system.}
  \label{fig:neighboring_same_type_water_phobic}
 \end{center}
\end{figure}

Fig.~\ref{fig:neighboring_same_type_water_phobic} shows the time
evolution of mean numbers of neighboring molecules of the same types
(water and hydrophobic monomer)---calculated as $\langle
|\{l(\VECTOR{i}) \mid \MType_{l(\VECTOR{i})} = X\}|
\rangle_{\VECTOR{i} \ \mathrm{s.t.}\ \MType_\VECTOR{i}=X}$ for
$X=\WATER, \PHOBIC$, respectively---in the same simulation.
The mean neighboring molecules of the same types start from the initial
random configuration (where the value $ \sim 0.25 \times 6 = 1.5$ for
both water and hydrophobic monomers) and increase relatively slowly,
taking a few million steps to reach the equilibrium state.  This result is
consistent with observations of the physico-chemical molecular
aggregation process, where small clusters are quickly formed, but as the
size grows, their mobility decreases and integration into larger
clusters requires more time.

\subsection{Amphiphilic polymers in a polar environment}\label{subsec:polymer_dynamics}

\begin{figure}[th]
 \begin{center}
 \includegraphics[width=1.0\textwidth]{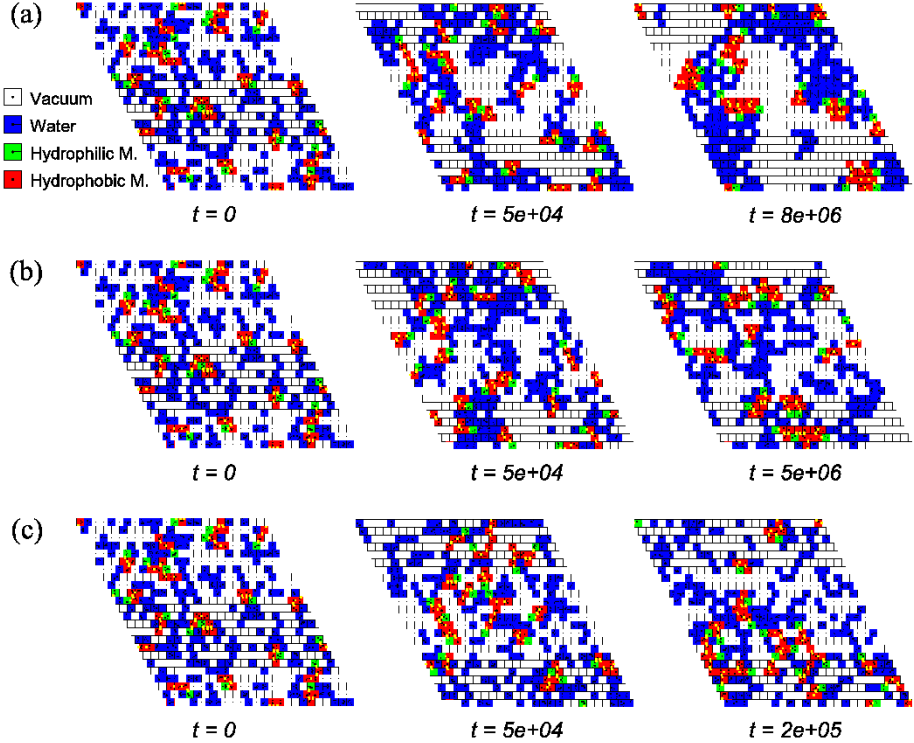}
  \caption{Snapshots of the molecular layer in three simulations of
  water--amphiphilic polymer systems with different initial
  distributions of heat particles (``temperature''): 
  (a) low-, (b) moderate-, and (c) high-temperature conditions.}
  \label{fig:molecular_snapshots_polymer_water}
 \end{center}
\end{figure}

Fig.~\ref{fig:molecular_snapshots_polymer_water} shows 
snapshots of the molecular layer in three simulations of 
amphiphilic tetramers (each consisting of three hydrophilic monomers plus
one hydrophobic head monomer, see Fig.~\ref{fig:illustration_polymer})
in solvent water, with different settings for the initial distribution of
heat particles. 

In the simulation of
Fig.~\ref{fig:molecular_snapshots_polymer_water}(a), initially, there are
no heat particles, representing the ``low-temperature'' condition (mean
total energy per cell $e_{\mathrm{total}} = 0.95$).
%
%
Although the low-temperature condition is the same as the one adopted in
the simulation of Fig.~\ref{fig:molecular_aggregation_water_phobic},
this simulation requires a longer relaxation time, because the polymers'
mobility is lower than that of the monomers (mainly because of the bond
maintenance condition
(\ref{eq:pair_interaction_bond_maintenance_condition})).
Starting from the initial condition where the tetramers are
homogeneously distributed, they aggregate into micelle-like
structures, their hydrophilic heads staying in contact with water and their
hydrophobic tails trying to cluster. 
The micelle-like structures are an elementary example of 
\STRESS{higher-order} structures\cite{Rasmussen2001}, with emergent properties
such as integrity and even lower mobility. 

Fig.~\ref{fig:molecular_snapshots_polymer_water}(b) corresponds to the
``moderate-temperature'' condition, where the initial 
$\Heat_{\VECTOR{i},l}$ is given randomly from $[0,1]$ with 
$\langle \Heat_{\VECTOR{i},l} \rangle_{\VECTOR{i}}=1$ for $l \in L$ (mean
total energy per cell $e_{\mathrm{total}} = 3.93$).
Compared with the low-temperature condition, while the sizes of the
organized micelle-like structures and water aggregates become smaller,
their motion becomes faster.

Fig.~\ref{fig:molecular_snapshots_polymer_water}(c) corresponds to the
``high-temperature'' condition, where the initial 
$\Heat_{\VECTOR{i},l}$ is given randomly from $[0,4]$ with 
$\langle \Heat_{\VECTOR{i},l} \rangle_{\VECTOR{i}}=2$ for $l \in L$ 
(mean total energy per cell $e_{\mathrm{total}} = 13.01$).
In this condition, the molecular motion becomes even faster and no
distinct self-organization is observed.

\begin{figure}[th]
 \begin{center}
  \includegraphics[width=1.0\textwidth]{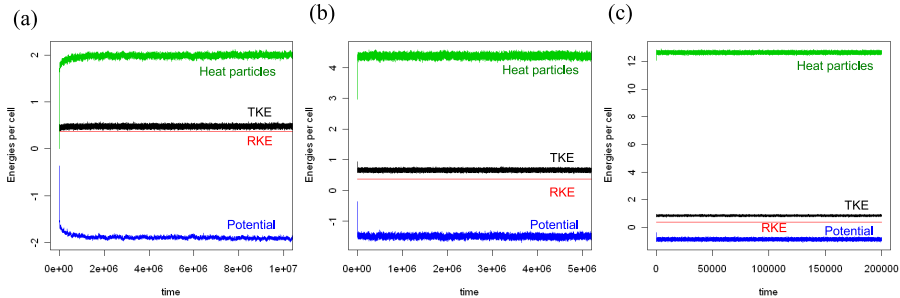}
 \caption{Time evolution of mean values of TKE, RKE, potential energy, and
 heat particles per cell in a simulation of water--amphiphilic polymer
 systems in the different temperature conditions: (a) low-,
 (b) moderate-, and (c) high-temperature conditions.}
 \label{fig:energies_transitions_polymer_water_different_temperature}
 \end{center}
\end{figure}

\begin{figure}[th]
 \begin{center}
  \includegraphics[width=1.0\textwidth]{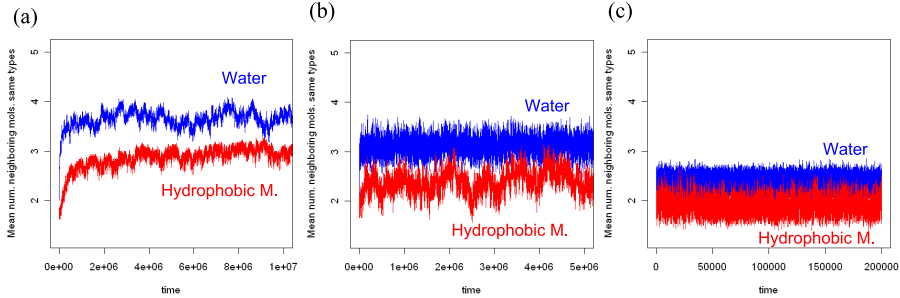}
 \caption{Time evolution of mean numbers of neighboring molecules of the
 same types (water and hydrophobic monomer) in the simulations of
 water--amphiphilic polymer systems with the different temperature
 conditions: (a) low-, (b) moderate-, and (c) high-temperature conditions.}
 \label{fig:neighboring_same_type_polymer_water_different_temperature}
 \end{center}
\end{figure}

Fig.~\ref{fig:energies_transitions_polymer_water_different_temperature}
shows the time evolution of mean energies per cell,
and
Fig.~\ref{fig:neighboring_same_type_polymer_water_different_temperature}
shows the time evolution of mean numbers of neighboring molecules of the same
types (water and hydrophobic monomer)
in the three abovementioned simulations with the different temperature
conditions.  Note the difference in the time scales.
These results indicate the temperature dependency of the molecular process.
That is, at lower temperature, polymers aggregate into larger structures;
however, the formation process takes a longer time.  On the other hand, at
higher temperature, large structures cannot be maintained while the
motion of polymers gets faster.
This kind of temperature dependency is derived (rather than being
presupposed) in a precise manner only from dynamical models with
reversibility and energy conservation.
Fig.~\ref{fig:neighboring_same_type_polymer_water_different_temperature}
(especially (a)) also shows that the aggregation of polymers is slower
than the clustering of water.

\subsection{Phase separation dynamics in ternary mixtures}\label{subsec:ternary_mixtures}

To compare in more detail the behavior of our RLMA with experimental
observations and other models (especially the Larson-type ones), we
conducted simulation of ternary mixtures of water, hydrophobic monomers,
and amphiphilic polymers, and analyzed the phase separation dynamics
with different concentration and temperature (total energy) settings.

Theories as well as successful models have shown that 
the phase separation or domain growth dynamics generally obeys dynamic
scaling\cite{Binder1974,Koch1983,Amar1988,Liverpool1996,Bernardes1996},
where domain structure remains statistically invariant in time under
rescaling by the characteristic length scale $L$, and $L$ grows as a
function of time following the asymptotic power law, $L(t) \sim
t^{1/z}$.  The theories typically suggest $z=1/3$, though the value can
differ depending on the stages of phase separation.


To check if the RLMA realizes the dynamic scaling behavior, 
we investigated the evolution of \STRESS{mean cluster radius}.
A molecule of type $X$ (in this case, $X$ can be water $\WATER$,
hydrophobic monomer $\PHOBIC$, or amphiphile $\AMPHIPHILE$) belongs to a
cluster of type $X$ if any of its nearest neighbor are of the same type
and are already counted as part of the cluster.
%
Using the cluster distribution $\{n_{X}(s)\}_s$, where $n_{X}(s)$ is the
number of type $X$ clusters with size $s$, the mean cluster size
$\chi_{X}$ of type $X$ is estimated as
\begin{equation}
 \label{eq:mean_cluster_size}
  \chi_{X} = \sum_{s=1}^{s_{\mathrm{max},X}-1} s^2 n_{X}(s) \left/ \ 
  \sum_{s=1}^{s_{\mathrm{max},X}} s n_{X}(s), \right.
\end{equation}
where $s_{\mathrm{max},X}$ is the largest cluster of type $X$.
In two dimensions, the mean cluster radius of type $X$ is estimated as
$R_X \sim \chi_{X}^{1/2}$.

\begin{figure}[th]
 \begin{center}
  \includegraphics[width=0.75\textwidth]{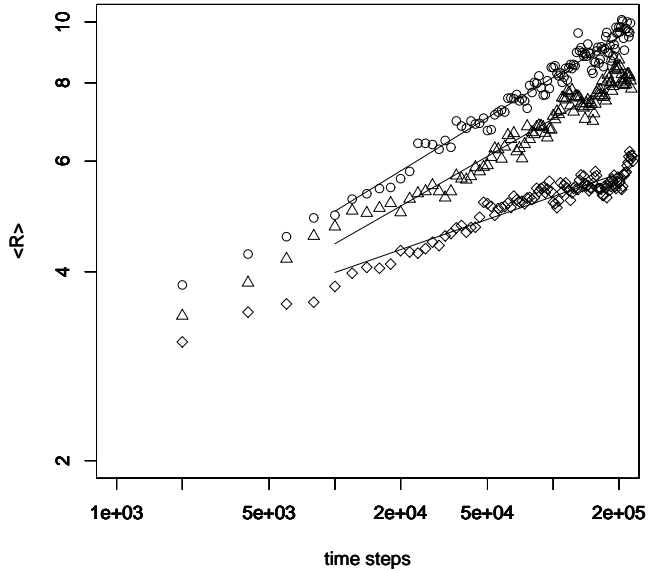}
 \caption{Time evolution of the averaged mean cluster radius $\langle R
 \rangle$, with different molecular ratios.  $\circ$ represents
 $(\phi_{\WATER}, \phi_{\PHOBIC}, \phi_{\AMPHIPHILE}) = (0.288, 0.288,
 0.024)$, $\triangle$ represents $(\phi_{\WATER}, \phi_{\PHOBIC},
 \phi_{\AMPHIPHILE}) = (0.27, 0.27, 0.06)$, and $\diamond$ represents
 $(\phi_{\WATER}, \phi_{\PHOBIC}, \phi_{\AMPHIPHILE}) = (0.24, 0.24,
 0.12)$. The lines show power law relations with estimated scaling
 exponents $1/z$ for their slopes.}
 \label{fig:dynamics_average_cluster_radius}
 \end{center}
\end{figure}

Fig.~\ref{fig:dynamics_average_cluster_radius} shows the evolution of
averaged mean cluster radius $\langle R \rangle =
[\frac{1}{2}(\chi_{\WATER}) + \chi_{\PHOBIC})]^{1/2}$, with different
concentration ratios of water $\phi_{\WATER}$, hydrophobic monomers
$\phi_{\PHOBIC}$, and amphiphilic polymers $\phi_{A}$ (the ratio of
vacuum $\phi_{\VACUUM}=0.4$ is common), in a temperature setting (mean
total energy $e_{\mathrm{total}} = 1.18$).  Power law behavior is
observed for all the concentration ratios $(\phi_{\WATER},
\phi_{\PHOBIC}, \phi_{\AMPHIPHILE}) = (0.288, 0.288, 0.024)$, $(0.27,
0.27, 0.06)$, and $(0.24, 0.24, 0.12)$.
For each setting, the scaling exponent $1/z$ is estimated as $0.213 \pm
0.004$, $0.199\pm 0.005$, and $0.121\pm 0.004$, by fitting the data
within time region $[10000, 200000]$ into $\langle R \rangle(t) \sim
t^{1/z}$.
These estimated values, especially the former two ($1/z \sim 0.2$) are
similar to the ones obtained by simpler Ising spin models for binary
systems \cite{Amar1988,Glotzer1994,Liverpool1996}, but smaller than the
theoretical value $z=1/3$ which is also obtained in
Ref.~\cite{Bernardes1996} by adding small amount of amphiphile into
binary mixture, like in this simulation.
The small values are possibly because the asymptotic late stage is not
reached due to the small size of the system, 
but other possible reasons can be also suggested:
(i) the existence of hydrodynamics, which is supposed to be absent in
the dynamic scaling hypothesis\cite{Binder1974} as well as the
Larson-type model in Ref.~\cite{Bernardes1996},
(ii) the constant-energy condition, where the temperature increases as
the clustering proceeds and the dynamic exponent becomes smaller
compared to the isothermal condition\cite{Koch1983}, while the latter is
usually used in the Larson-type models and other Monte Carlo methods,
and
(iii) other conservation laws, which also work to decrease the exponent
\cite{Liverpool1996}.
The decrease in the growth rate of $\langle R \rangle$ accompanying the
increase of concentration $\phi_{\AMPHIPHILE}$ of amphiphile is
consistent with the result in Ref.~\cite{Bernardes1996}.


For different temperature settings we also calculated the
\STRESS{equal-time structure factors} $S_{X}(k,t)$, which is the Fourier
transform of the \STRESS{equal-time pair correlation function}, defined as
\begin{equation}
 \label{eq:structure_factor}
 S_{X}(k,t) = \int C_{XX}(\vec{r}, t) e^{i \vec{k} \cdot \vec{r}} d\vec{r}, 
\end{equation}
\begin{equation}
 \label{eq:pair_correlation}
  C_{XX}(\vec{r}, t) = 
  \langle {\delta \rho_{X}}(\vec{r}, t)\;{\delta \rho_{X}}(\vec{0}, t) \rangle.
\end{equation}
The equal-time pair correlation $C_{XX}(\vec{r}, t)$ of type $X$ is
calculated by drawing shells of radius $r$ and $r+1$ around each
molecule of type $X$, counting the number of the same type molecules
between the shells, and finally normalizing by dividing by $r$.

\begin{figure}[th]
 \begin{center}
  \begin{minipage}{0.3\textwidth}
   (a)
   \includegraphics[width=1.0\textwidth]{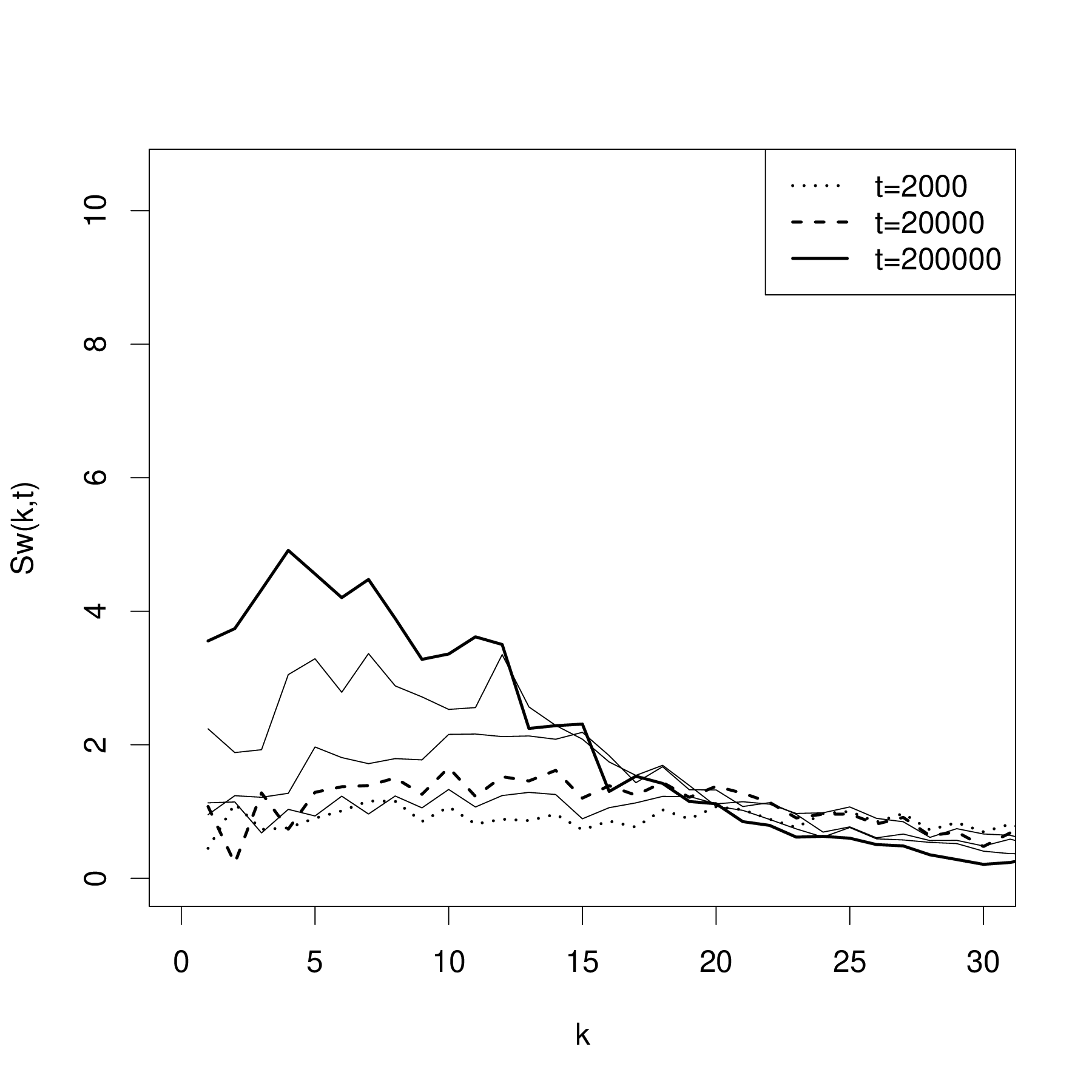}
  \end{minipage}\hfill
  \begin{minipage}{0.3\textwidth}
   (b)
   \includegraphics[width=1.0\textwidth]{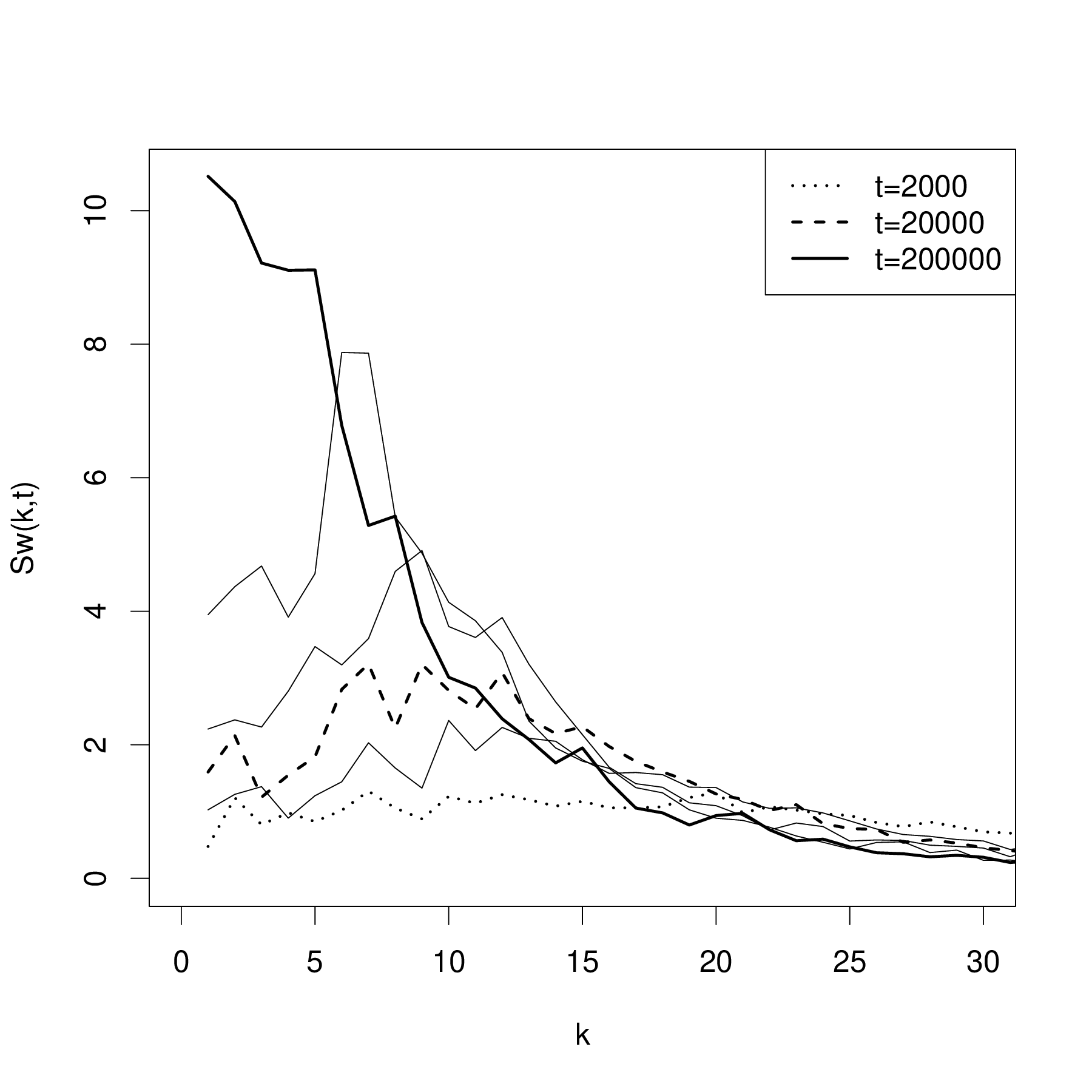}
  \end{minipage}\hfill
  \begin{minipage}{0.3\textwidth}
   (c)
   \includegraphics[width=1.0\textwidth]{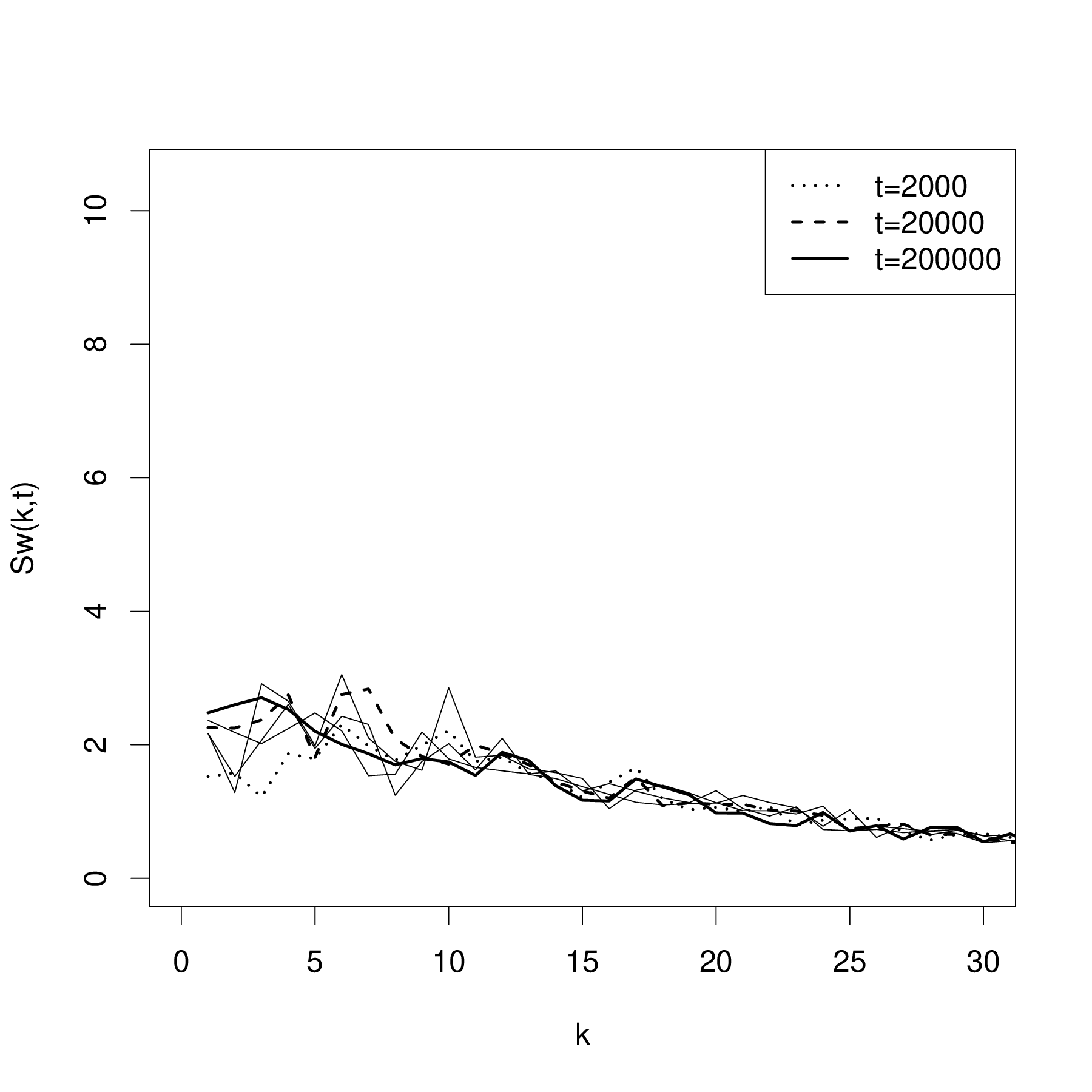}
  \end{minipage}\hfill
 \caption{The equal-time water--water structure factor for a system with
  $(\phi_{\WATER}, \phi_{\PHOBIC}, \phi_{\AMPHIPHILE}) = (0.288, 0.288,
  0.024)$ and different temperature (total energy) settings: (a)
  $e_{\mathrm{total}} = 0.37$, (b) $e_{\mathrm{total}} = 1.18$, (c)
  $e_{\mathrm{total}} = 5.63$.  $S_{\WATER}(k,t)$ at times $t = 2, 10, 20,
  50, 100, 200 \times 10^3$ are shown.}
  \label{fig:structure_factors_for_different_temperatures}
 \end{center}
\end{figure}

Fig.~\ref{fig:structure_factors_for_different_temperatures} shows the
equal-time water--water structure factor $S_{\WATER}(k,t)$ for a system
with $(\phi_{\WATER}, \phi_{\PHOBIC}, \phi_{\AMPHIPHILE}) = (0.288,
0.288, 0.024)$ and different temperature (total energy) settings.
In the lower temperature settings as 
Fig.~\ref{fig:structure_factors_for_different_temperatures}(a) and (b)
(mean total energy $e_{\mathrm{total}}=0.37$ and
$e_{\mathrm{total}}=1.18$, respectively), 
the structure factor develops a peak at nonzero wave number that grows
in time and the position of the peak moves to lower $k$ as $t$
increases. 
The peak at $k \sim 1$ indicates that the system is approaching to a global
separation into spanning networks.
The growth rate of the peak is higher for the ``warm'' condition
$e_{\mathrm{total}}=1.18$ than the ``cool'' condition
$e_{\mathrm{total}}=0.37$.
Above a critical total energy (that corresponds to the \STRESS{demixing
temperature}), as
Fig.~\ref{fig:structure_factors_for_different_temperatures}(c)
($e_{\mathrm{total}}=5.63$), the structure factor does not show any
structure.  These are again in good agreement with the result in
Ref.~\cite{Bernardes1996}, as well as experiment\cite{Mallamace1995}.

\section{Conclusion}\label{sec:discussion}


In this paper, we described the construction of RLMA, which simulates
physico-chemical interaction of molecules and their self-organization
process.
The definition of the model has shown how to eliminate the
irreversibility in the original LMA using several techniques to
construct reversible CA. 
Simulation results of RLMA dynamics have demonstrated that 
the RLMA can deal with broader situations, with the original LMA's
self-organization results as special cases.
%
The results also showed that the RLMA reproduces qualitatively
consistent results with the traditional Larson-type models.

Although several reversible CA models have been proposed to
simulate self-organization processes\cite{Creutz1986,D'Souza1999}, to
our knowledge, this is the first deterministic CA model that simulates
self-organization of \STRESS{higher-order} structures,
while satisfying strict reversibility.

Reversibility and conservation laws of the model enable precise
application and validation of the various methods in both equilibrium
and nonequilibrium statistical mechanics.
Reversibility also enables rigorous tracking of the information flow
driven by the dynamics, with no veiled sources or sinks. 
Therefore, the model will be preferable in analyzing the
self-organization and dynamics of multiple levels of structures from a
information-theoretic viewpoint (\textit{e.g.},
Ref.~\cite{McGregor2005}), as well as from the physically grounded
viewpoint.
%
%

This study focused on the process of molecular assembly. However, the
model can be extended to incorporate chemical reactions and catalytic
effects, by introducing more types of molecules and setting proper
values of excitation energies for the reactions, with their modulation
in the existence of neighboring catalytic molecules.
We are currently working on the construction of such a reversible and
thermodynamically consistent model that realizes
``protocells''\cite{Ono2005} with self-maintenance of compartment
structures, metabolism, and self-reproduction.

\section*{Acknowledgments}

This study was in part supported by
``Symbiotic Information Technology Research Project'' of Tokyo
University of Agriculture and Technology, and also by the Grant-in-Aid
for ``Scientific Research on Priority Areas (Area No. 454)'' from the
Japanese Ministry of Education, Culture, Sports, Science and Technology.

\appendix

\section{Alternative implementation of molecular rotation}\label{appsec:alternative_rotation}

Here, we present an alternative rotation rule, which 
enables the stationary state $\RKE=0$ for the polar molecules.
Using the same interleaved scheme of
Fig.~\ref{fig:molecular_rotation_time_scheme}, the rotational
permutation $\PermRot$ of (\ref{eq:rotational_update}) is replaced by
the new conditional permutation $\AltPermRot$ defined below.


First, consider that a polar molecule at $\VECTOR{i}$ is rotating, i.e., 
$\RKE_\VECTOR{i} \neq 0$.  Then, $\AltPermRot$ maintains the rotation and
changes the molecule's orientation by
$\DirRotation^{\SIGN(\RKE_\VECTOR{i})}$ %
\footnote{The sign function, $\SIGN(\RKE_\VECTOR{i}) = +1 \mbox { if }
\RKE_\VECTOR{i}>0, -1 \mbox{ if } \RKE_\VECTOR{i}<0$, indicates the direction
of rotation.} 
if the magnitude of
$\RKE_\VECTOR{i}$ is more than sufficient to compensate for the change in
potential induced by the rotation; else, it inverts the direction of
rotation if the magnitude of $\RKE_\VECTOR{i}$ is not large enough.
If the magnitude of $\RKE_\VECTOR{i}$ is just sufficient to compensate for
the potential change, $\AltPermRot$ either executes the rotation and brings
the molecule to the stationary state, or inverts the direction of
rotation, depending on some conditions to avoid irreversibility:
\begin{equation}
 \label{eq:alt_rotation_update_nonzero_RKE}
   \AltPermRot
   \left(
    \begin{array}{c}
     \MOri_\VECTOR{i}\\
      \RKE_\VECTOR{i}\\
    \end{array}   
   \right)
   = 
   \left\{
    \begin{array}{l}
     \left(
      \begin{array}{c}
       \DirRotation^{\SIGN(\RKE_\VECTOR{i})}(\MOri_\VECTOR{i})\\
       \RKE_\VECTOR{i} - \SIGN(\RKE_\VECTOR{i})\PotentialChangeByRotation{\MOri_\VECTOR{i}}{\SIGN(\RKE_\VECTOR{i})}^{\VECTOR{i}}
      \end{array}
     \right)
     \quad \mbox{ if }
     \RKE_\VECTOR{i} \neq 0 \mbox{ and }
     \\
     \qquad
      \begin{array}{ccl}
       \left\{
	\begin{array}{l}
	 |\RKE_\VECTOR{i}| > \PotentialChangeByRotation{\MOri_\VECTOR{i}}{\SIGN(\RKE_\VECTOR{i})}^{\VECTOR{i}}, \mbox{ or}\\
	 |\RKE_\VECTOR{i}| = \PotentialChangeByRotation{\MOri_\VECTOR{i}}{\SIGN(\RKE_\VECTOR{i})}^{\VECTOR{i}} 
	  \\ \qquad
	 \wedge 
	  \PotentialChangeByRotation{\DirRotation^{\SIGN(\RKE_\VECTOR{i})}(\MOri_\VECTOR{i})}{\SIGN(\RKE_\VECTOR{i})}^{\VECTOR{i}} \geq 0
	 , \mbox{ or}\\
	 |\RKE_\VECTOR{i}| = \PotentialChangeByRotation{\MOri_\VECTOR{i}}{\SIGN(\RKE_\VECTOR{i})}^{\VECTOR{i}} 
	  \\ \qquad
	  \wedge 
	  \PotentialChangeByRotation{\DirRotation^{\SIGN(\RKE_\VECTOR{i})}(\MOri_\VECTOR{i})}{\SIGN(\RKE_\VECTOR{i})}^{\VECTOR{i}} < 0
	  \\ \qquad\qquad
	  \wedge \PARITY(\PDir_\VECTOR{i}) \neq \SIGN(\RKE_\VECTOR{i})
	  ,\\
	\end{array}
       \right\} 
      \end{array}   
      \\
     \left(
      \begin{array}{c}
       \MOri_\VECTOR{i}\\
       -\RKE_\VECTOR{i}
      \end{array}
     \right)
     \quad \mbox{ if }
     \RKE_\VECTOR{i} \neq 0 \mbox{ and }
     \\
     \qquad
     \left\{
      \begin{array}{l}
       |\RKE_\VECTOR{i}| < \PotentialChangeByRotation{\MOri_\VECTOR{i}}{\SIGN(\RKE_\VECTOR{i})}^{\VECTOR{i}}, \mbox{ or}\\
       |\RKE_\VECTOR{i}| = \PotentialChangeByRotation{\MOri_\VECTOR{i}}{\SIGN(\RKE_\VECTOR{i})}^{\VECTOR{i}} 
	\\ \qquad
	\wedge 
	\PotentialChangeByRotation{\DirRotation^{\SIGN(\RKE_\VECTOR{i})}(\MOri_\VECTOR{i})}{\SIGN(\RKE_\VECTOR{i})}^{\VECTOR{i}} < 0
	\\ \qquad\qquad
	\wedge \PARITY(\PDir_\VECTOR{i}) = \SIGN(\RKE_\VECTOR{i}).
	\\
      \end{array}
     \right\} 
    \end{array}
   \right.
\end{equation}

Next, consider that a polar molecule at $\VECTOR{i}$ is in the
stationary state, i.e., $\RKE_\VECTOR{i} = 0$.  Then, $\AltPermRot$
starts the rotation if changing the molecule's orientation by one of the
directions $\DirRotation^{\pm 1}$ induces a negative potential change.
If rotations in both of the directions $\DirRotation^{\pm 1}$ induce
negative potential changes, $\AltPermRot$ starts the rotation according
to the preferential direction.
On the other hand, if the molecule's orientation is at a local potential
minimum, the molecule maintains its stationary state:
\begin{equation}
 \label{eq:alt_rotation_update_zero_RKE}
   \AltPermRot
   \left(
    \begin{array}{c}
     \MOri_\VECTOR{i}\\
      \RKE_\VECTOR{i}\\
    \end{array}   
   \right)
   = 
   \left\{
    \begin{array}{l}
     \left(
      \begin{array}{c}
       \DirRotation^{\pm 1}(\MOri_\VECTOR{i})\\
       \mp\PotentialChangeByRotation{\MOri_\VECTOR{i}}{\pm 1}^{\VECTOR{i}}
      \end{array}
     \right)
   \quad \mbox{ if }
   \RKE_\VECTOR{i} = 0 \mbox{ and}
   \\
     \qquad
   \left\{
   \begin{array}{l}
    \PotentialChangeByRotation{\MOri_\VECTOR{i}}{\pm 1}^{\VECTOR{i}} < 0
     \\ \qquad
     \wedge
    \PotentialChangeByRotation{\MOri_\VECTOR{i}}{\mp 1}^{\VECTOR{i}} \geq 0
    , \mbox{ or}
    \\
    \PotentialChangeByRotation{\MOri_\VECTOR{i}}{+1}^{\VECTOR{i}} < 0
     \\ \qquad
     \wedge
     \PotentialChangeByRotation{\MOri_\VECTOR{i}}{-1}^{\VECTOR{i}} < 0
     \wedge
     \PARITY(\PDir_\VECTOR{i}) = \pm 1,
   \end{array}
   \right\} 
   \\
     \left(
      \begin{array}{c}
       \MOri_\VECTOR{i}\\
       0
      \end{array}     
     \right)
   \quad \mbox{ if } 
   \RKE_\VECTOR{i} = 0 \mbox{ and }
   \\ \quad\qquad
    \PotentialChangeByRotation{\MOri_\VECTOR{i}}{+1}^{\VECTOR{i}} \geq 0
     \wedge
    \PotentialChangeByRotation{\MOri_\VECTOR{i}}{-1}^{\VECTOR{i}} \geq 0.
    \end{array}
  \right.
\end{equation}

Here 
\begin{equation}
 \label{eq:sum_def_potential_change_by_rotation}
 \PotentialChangeByRotation{k}{n}^\VECTOR{i}
 = \sum_{l \in L} \PotentialChangeByRotation{k}{n}^{\VECTOR{i}, l(\VECTOR{i})}
\end{equation}
represents the total potential change at $\VECTOR{i}$ that occurs when the
orientation of the molecule at $\VECTOR{i}$ is changed from $k$ to 
$\DirRotation^{n}(k)$ 
(see the notation (\ref{eq:def_potential_change_by_rotation}), too).

The main point is that, in $\AltPermRot$, change in RKE and not heat
compensates for the change in potential.
Therefore, the RKE layer can work as another energy storage.
Recall that in the rotation rule (\ref{eq:rotational_update}),
RKE works just as ``second-order'' signals to preserve
reversibility; hence, it cannot change to $0$.
It should also be noted that the parity (\ref{eq:def_parity_pd}) of the
preferential direction is utilized to avoid non-uniqueness of the
$\AltPermRot$'s pre-images, which could be derived from unstable fixed
points (stationary states at orientations of local maximum potential).

One drawback of this alternative rotation rule is that the value of RKE
is unbounded in principle; thus, the model is not a CA in the strict sense.
In practice, however, due to the energy conservation
(\ref{eq:energy_conservation_law}), limitless divergence of RKE cannot
occur unless an infinite amount of energy is injected into a finite
region.

\begin{figure}[th]
 \begin{center}
 \includegraphics[width=1.0\textwidth]{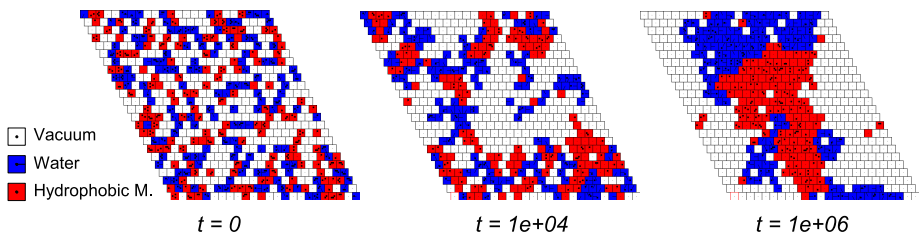}
  \caption{Snapshots of the molecular layer in a simulation of
  a water--hydrophobic monomer system, using the alternative rotation rule
  $\AltPermRot$. Clustering and phase separation occur in a
  similar manner to those shown in Fig.~\ref{fig:molecular_aggregation_water_phobic}.}
  \label{fig:alt_molecular_aggregation_water_phobic}
 \end{center}
\end{figure}

\begin{figure}[th]
 \begin{center}
 \includegraphics[width=0.45\textwidth]{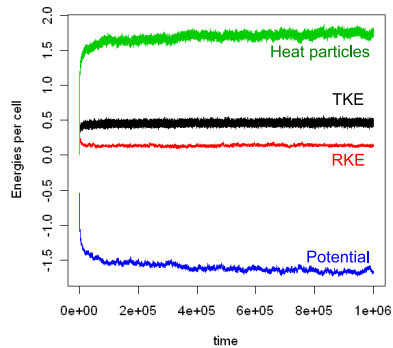}
 \caption{Time evolution of mean values of energies per cell in a simulation
 of a water--hydrophobic monomer system, using $\AltPermRot$. The mean
 absolute value of RKE $\langle |\RKE_\VECTOR{i}| \rangle_{\VECTOR{i}} $
 fluctuates; this is in contrast to
 Fig.~\ref{fig:energies_transition_water_phobic} where the value is
 constant. }
  \label{fig:alt_energies_transition_water_phobic}
 \end{center}
\end{figure}

\begin{figure}[th]
 \begin{center}
 \includegraphics[width=0.45\textwidth]{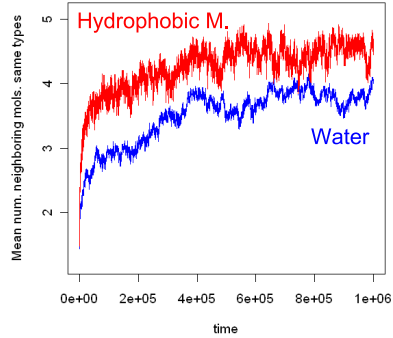}
  \caption{Time evolution of mean numbers of neighboring molecules of the
  same types in the simulation of a water--hydrophobic monomer system,
  using $\AltPermRot$. The evolution is similar to
  that shown in Fig.~\ref{fig:neighboring_same_type_water_phobic}.}
  \label{fig:alt_neighboring_same_type_water_phobic}
 \end{center}
\end{figure}

The RLMA model with the alternative rotation permutation $\AltPermRot$
shows qualitatively similar behavior.
Fig.~\ref{fig:alt_molecular_aggregation_water_phobic}
shows snapshots of the molecular layer,
Fig.~\ref{fig:alt_energies_transition_water_phobic}
shows the time evolution of mean energies per cell, 
and Fig.~\ref{fig:alt_neighboring_same_type_water_phobic}
shows the time evolution of mean numbers of neighboring molecules 
of the same types
in a simulation of a water--hydrophobic monomer system, with the same
initial configuration as that in 
section~\ref{subsec:water_phobic_clustering}.

\end{document}